\documentclass[twocolumn,twocolappendix]{aastex63}
\usepackage[T1]{fontenc}
\usepackage{ae,aecompl}
\usepackage{mathtools,tabu}
\usepackage{amsmath,amssymb}
\usepackage{upgreek} 

\newcommand{\avg}[1]{\left\langle{#1}\right\rangle}
\newcommand{\comment}[1]{}
\newcommand{\beq}{\begin{equation}}
\newcommand{\eeq}{\end{equation}}
\newcommand{\beqa}{\begin{equation}\begin{aligned}}
\newcommand{\eeqa}{\end{aligned}\end{equation}}
\newcommand{\bit}{\begin{itemize}}
\newcommand{\eit}{\end{itemize}}

\newcommand{\hiGpc}{h^{-1} \rm Gpc}
\newcommand{\hiMpc}{h^{-1} \rm Mpc}

\newcommand{\hiMsun}{h^{-1} M_\odot}

\newcommand{\DS}{\Delta\Sigma}

\newcommand{\psrc}{p_{\rm src}}

\newcommand{\rp}{r_{\rm p}}

\newcommand{\dd}{{\rm d}}
\newcommand{\xihm}{\xi_{\rm hm}}

\newcommand{\OmegaM}{\Omega_{\rm M}}

\newcommand{\lnsigeight}{\ln\sigma_8}

\newcommand{\lnM}{\ln M}
\newcommand{\Mobs}{M_{\rm obs}}
\newcommand{\Mobsth}{M_{\rm obs}^{\rm th}}
\newcommand{\lnMobs}{{\ln M_{\rm obs}}}
\newcommand{\lnMobsth}{\ln M_{\rm obs}^{\rm th}}

\newcommand{\Meqth}{M_{\rm eq}^{\rm th}}

\newcommand{\Msplit}{M_{\rm split}}

\newcommand{\Mtrue}{M_{\rm true}}

\newcommand{\sigmalnM}{\sigma_{\ln M}}

\newcommand{\zlens}{z_{\rm lens}}
\newcommand{\zsrc}{z_{\rm src}}

\newcommand{\avgn}{\langle n \rangle}

\newcommand{\xicm}{\xi_{\rm cm}}

\newcommand{\rpmin}{{\rm r_p^{min}}}
\newcommand{\rpmax}{{\rm r_p^{max}}}

\newcommand{\nsrc}{{n_{\rm src}}}

\shorttitle{Cluster cosmology forecast}
\shortauthors{Wu et al.}
\begin{document}

\title{Cosmology with galaxy cluster weak lensing: statistical limits and experimental design}

\author[0000-0002-7904-1707]{Hao-Yi Wu}
\affiliation{Department of Physics, Boise State University, Boise, ID 83725, USA}
\affiliation{Department of Physics, Department of Astronomy, and Center for Cosmology and Astro-Particle Physics, \\ The Ohio State University, Columbus, OH 43210, USA}

\author[0000-0001-7775-7261]{David H. Weinberg}
\affiliation{Department of Physics, Department of Astronomy, and Center for Cosmology and Astro-Particle Physics, \\ The Ohio State University, Columbus, OH 43210, USA}
\affiliation{Institute for Advanced Study, Princeton, NJ 08540, USA}

\author[0000-0003-1420-527X]{Andr\'{e}s N. Salcedo}
\affiliation{Department of Physics, Department of Astronomy, and Center for Cosmology and Astro-Particle Physics, \\ The Ohio State University, Columbus, OH 43210, USA}

\author[0000-0003-3175-2291]{Benjamin D. Wibking}
\affiliation{Australian National University, Mount Stromlo Observatory, Cotter Road, Weston Creek, ACT 2611, Australia }

\begin{abstract}

We forecast constraints on the amplitude of matter clustering $\sigma_8(z)$ achievable with the combination of cluster weak lensing and number counts, in current and next-generation weak lensing surveys. We advocate for an approach, analogous to galaxy--galaxy lensing, in which the observables in each redshift bin are the mean number counts and the mean weak lensing profile of clusters above a mass proxy threshold. The primary astrophysical nuisance parameter is the logarithmic scatter $\sigmalnM$ between the mass proxy and true mass near the threshold. For surveys similar to the Dark Energy Survey (DES), the {\rm Roman Space Telescope} High Latitude Survey (HLS), and the Rubin Observatory Legacy Survey of Space and Time (LSST), we forecast aggregate precision on $\sigma_8$ of 0.26\%, 0.24\%, and 0.10\%, respectively, {\em if} the mass--observable scatter is known externally to $\Delta\sigmalnM \leq 0.01$. These constraints would be degraded by about 20\% for $\Delta\sigmalnM = 0.05$ in the case of DES or HLS and for $\Delta\sigmalnM = 0.016$ for LSST. A 1 month observing program with {\rm Roman Space Telescope} targeting $\sim 2500$ massive clusters could achieve a $\sim 0.5\%$ constraint on $\sigma_8(z=0.7)$ on its own, or a $\sim 0.33\%$ constraint in combination with the HLS. Realizing the constraining power of clusters requires accurate knowledge of the mass--observable relation and stringent control of  systematics. We provide analytic approximations to our numerical results that allow for easy scaling to other survey assumptions or other methods of cluster mass estimation.

\end{abstract}


\section{Introduction}

The masses of galaxy clusters have long been recognized as a powerful probe of matter clustering in the low redshift universe (\citealt{Evrard89,Peebles89}; see \citealt{Allen11} for a review).  In an influential study combining analytic approximations and numerical simulations, \cite{White93} showed that predicted cluster masses depend most strongly on the matter density parameter $\OmegaM$ and the power spectrum amplitude $\sigma_8$.  The parameter combination best constrained by low redshift cluster masses is approximately $\sigma_8\OmegaM^{0.5}$, and \cite{White93} used observationally estimated masses of clusters to conclude that either $\OmegaM$ was substantially lower than unity or the matter $\sigma_8 \approx 0.6$ was substantially lower than the corresponding value for galaxies.  Many subsequent studies have focused on the evolution of cluster abundances, with redshift evolution providing a further sensitive probe of cosmology \citep[e.g.,][]{Holder01,Haiman01,Miller01,KravtsovBorgani12}.  The mass function of dark matter halos in co-moving coordinates tracks the growth rate of matter fluctuations, and the total number of clusters in a survey volume further depends on the cosmic expansion history.  Supernovae and baryon acoustic oscillations also afford precise measurements of the cosmic expansion history, but galaxy clusters remain one of the most powerful probes of matter clustering evolution, and thus a valuable tool for distinguishing between dark energy and modified gravity as explanations of cosmic acceleration (see \citealt{Weinberg13} for a comprehensive review).  In this paper, we focus on the structure growth constraints that can be derived from cluster populations in large area weak lensing surveys.

Many recent studies formulate the cluster cosmology problem as follows: (1) measure the number counts of clusters as a function of an observable mass proxy such as optical richness or X-ray luminosity; (2) use a dynamically sensitive observable such as velocity dispersion, X-ray temperature, Sunyaev--Zeldovich (SZ) decrement, or gravitational lensing to calibrate the mass--observable relation; and (3) combine these measurements to infer the cluster mass function and use it to constrain $\sigma_8$, $\OmegaM$, or (in combination with external data sets) other cosmological parameters  \citep[e.g.,][]{Vikhlinin09,Mantz10,Rozo10,Benson13,Planck13Cluster,Mantz14,Planck15Cluster,Bocquet19,DESY1CL}.  The second of these three steps is the most challenging, since one must reduce systematic errors in the cluster mass scale to the $\sim 1\%$ level or below to remain competitive with current and future constraints from other probes such as cosmic shear or galaxy redshift-space distortions. Even the best X-ray observables such as hydrostatic mass (e.g., \citealt{Vikhlinin09, Applegate16}) or the ``X-ray SZ parameter'' $Y_X$ \citep{Kravtsov06}, if predicted directly from theory, have likely systematics at the few percent level because of uncertain corrections for nonthermal pressure support and departures from hydrostatic equilibrium.

We regard weak gravitational lensing as the most promising cluster mass probe for precision cosmological investigations, in part because the relation between the projected mass distribution and the weak lensing signal can be predicted without theoretical uncertainty (provided one assumes that the theory of general relativity is correct), in part because weak lensing can probe the outskirts of clusters where baryonic effects on the mass distribution are less uncertain, and in part because the current and future generations of large weak lensing surveys provide ideal data sets for cluster weak lensing measurements.  One can still view weak lensing as calibrating the mass--observable relation, i.e., executing step (2) of the outline above. This is essentially the approach taken by the ``Weighing the Giants'' program \citep{vonderLinden14WtG} and several other recent studies \citep[e.g.,][]{Melchior17,Simet17,McClintock19}. Alternatively, one can view cluster weak lensing as analogous to galaxy--galaxy weak lensing \citep[e.g.,][]{McKay01,Mandelbaum06,Mandelbaum13}, where the basic observables are the number counts of the cluster sample and the excess surface density profile that governs the measured tangential shear.  The constraining power of optical cluster surveys is then determined by the accuracy of the weak lensing measurements.  Next generation optical imaging surveys will provide unprecedented accuracy for cluster lensing studies, using either clusters identified within the survey or clusters from extended X-ray or SZ surveys.

In this work, instead of the more common one-point statistics perspective (calibrating the mass--observable relation with weak lensing and inferring the halo mass function), we adopt a two-point statistics perspective, with cluster shear profiles as the main observable and the cluster number counts used to break degeneracy. This perspective can be generalized to encompass other two-point statistics such as cluster auto-correlations, the cluster--galaxy cross-correlation, and galaxy auto-correlation \citep[e.g.,][]{Chiu20,Salcedo20,To20a,To20b}.

We focus specifically on the ability of cluster weak lensing studies to constrain $\sigma_8(z)$, the amplitude of matter fluctuations in redshift bins, when the value of $\OmegaM$ and the shape of the matter power spectrum are known from external constraints.  While the degeneracy between $\sigma_8$ and $\OmegaM$ is likely to remain important in observational analyses, the ability to break this degeneracy depends strongly on one's assumptions about external data sets. It is thus simplest to characterize the constraining power of cluster weak lensing samples in terms of a single number at each redshift.

For testing some dark energy or modified gravity models, what matters is the ability to pin down the shape of $\sigma_8(z)$, tracking changes in the rate of structure growth at low redshift.  In other cases, the most stringent test comes from comparing the overall amplitude of low redshift matter clustering to the cosmic microwave background predictions for $\Lambda$CDM (cosmological constant and cold dark matter) or  an alternative dark energy model. If we assume $\Lambda$CDM and treat errors in different redshift bins as independent,  then we can combine the $\sigma_8(z)$ constraints from different redshift bins by summing the inverse square to obtain the aggregate uncertainty in $\sigma_8$ at $z=0$:  
\beq
\Delta\ln\sigma_8 = \left(\sum_i\left[\Delta\ln\sigma_8(z_i)\right]^{-2}\right)^{-1/2} \ .
\label{eq:aggregate} 
\eeq
The goal for observational analyses is to have this aggregate precision limited by the statistical uncertainties of the data set rather than by the systematic uncertainties.  For cluster weak lensing, the two main sources of statistical uncertainty are the finite size of the cluster sample and the weak lensing shape noise.

In the spirit of our galaxy--galaxy lensing analogy, we concentrate our analyses on constraints from the full population of clusters above some threshold in the observable mass proxy $\Mobs$.  In this treatment the ``nuisance parameters'' for inferring $\sigma_8(z)$ in a given redshift bin are the true mass $\Mtrue$ that corresponds to $\Mobs$ and the rms scatter $\sigmalnM$ between $\lnMobs$ and $\ln\Mtrue$ at this threshold.  We view this scatter as a quantity to be predicted from simulations or constrained from external data, such as high-quality X-ray observations of a subset of clusters in the weak lensing survey.  The value of $\sigma_8(z)$ and $\Mtrue$ can then be constrained using the mean weak lensing profile and number counts of clusters above the threshold.  We examine what knowledge of scatter is needed to avoid degrading statistical constraints on $\sigma_8(z)$.  In \citet{Salcedo20}, we show that combining cluster weak lensing with cluster--galaxy cross-correlations and galaxy auto-correlations, measurable in the same data used for cluster identification and weak lensing analysis, can constrain the scatter and break its degeneracy with the mass threshold and $\sigma_8$.

In principle, an analysis that uses multiple $\Mobs$ thresholds or divides clusters into $\Mobs$  bins could obtain additional cosmological constraining power by probing the shape of the halo mass function in addition to the mass and scatter at the $\Mobs$ threshold.  However, in our tests (see Appendix~\ref{app:bins}), we find that the gains from a multi-bin analysis are limited at best with the most general assumption for the mass--observable relation.  When considering multiple bins or multiple thresholds, one must introduce additional mass and scatter nuisance parameters at each boundary. Much if not all of the additional observational information goes to constraining these nuisance parameters rather than cosmology, and the modeling is more complicated because of the larger number of parameters and the need to accurately estimate the covariance of measurement errors and systematic errors across bins.  One can try to reduce the number of parameters by assuming, e.g., that the mass--observable relation and the scatter follow a power-law dependence on $M$ or $(1+z)$, as in \cite{Murata19} or \cite{Eifler20HLS}. However, this approach risks deriving spurious cosmological constraints from the cluster counts that are actually an artifact of the restrictive model.  We therefore focus our attention on the single-threshold analysis and advocate caution in using $\Mobs$ bins or multiple $\Mobs$ thresholds at a given redshift.

Although we focus on the $\sigma_8(z)$ constraints from the combination of cluster lensing and number counts, we also find it illuminating to compress the cluster lensing signal into the constraints on the mean cluster mass.  We will show that such data compression leads to slight loss of information, but it provides an easier way to assess the relative importance of the uncertainties of number counts (related to cluster sample size), uncertainties of  mean mass (related to lensing measurements), and the prior on scatter (from external data).  In this work we focus on the statistical uncertainties, but our formulae for parameter sensitivities are also applicable to systematic uncertainties or to alternative methods for calibrating cluster masses (e.g., multiwavelength, two-point statistics).

This paper is organized as follows. Section~\ref{sec:number} discusses the signal, noise, and parameter sensitivity of cluster number counts, while Section~\ref{sec:lensing} discusses those for cluster lensing.  Section~\ref{sec:constraints_per_volume} presents the general constraining power from combining cluster number counts and lensing, while Section~\ref{sec:surveys} incorporates specific survey assumptions.  In Section~\ref{sec:discusssions} we discuss  implications of our results, with particular attention to the control of mass--observable scatter and systematic errors needed to realize the statistical power of present and future surveys.  We summarize our results in Section~\ref{sec:summary}.

We adopt a flat $\Lambda$CDM cosmology with the following cosmological parameters:
$ w = -1$;
$\Omega_k$ = 0;
$\OmegaM$ = 0.314; 
$\Omega_{\Lambda}$ = 0.686; 
$h$ = 0.67; and
$\sigma_8$ = 0.83. 
These parameters follow those of the simulations from the Abacus Cosmos suite {\tt Abacus\_Cosmos\_720box\_planck} \citep{Garrison18}, which are used for calculating the small-scale lensing covariance matrices.
Throughout this work, 
cosmological distances are in co-moving units, 
and halo masses are based on a spherical density of 
200 times the background density ($M_{\rm 200m}$). 
We use the halo mass function and halo bias formulae provided by \cite{Tinker08MF,Tinker10Bias}.

\section{Galaxy cluster number counts and gravitational lensing}

\subsection{Cluster number counts}\label{sec:number}

When constructing a cluster sample, we set a threshold on the observed property $\Mobs$ (e.g., optical richness).  At a given mass $M$, let $\Mobs$ be the mass proxy re-scaled such that, 
\beq
\avg{\lnMobs | \lnM} = \lnM ~.
\eeq
The number density above the observable threshold $\Mobsth$ is given by
\beqa
\bar{n} 
& = \int_{\Mobsth}^{\infty}  d\Mobs \int_{0}^{\infty} dM  \frac{dn}{dM}  P(\Mobs | M) \\
&= \int d M \frac{dn}{dM}  \phi(\Mobsth | M)  \ .
\label{eq:counts}
\eeqa
The nuisance parameters are included in the selection function $\phi$.  If the mass--observable scatter is negligible, $\phi$ is described by a step function $H(M > \Mobsth)$.  In our calculation, we assume that $\phi$ is described by a complementary error function
\beq
\phi(\Mobsth | M ) = \frac{1}{2} {\rm erfc}
\bigg(
\frac{\lnMobsth - \lnM}{\sqrt{2\uppi {\sigma_{\lnM}^2}}} 
\bigg) ~,
\eeq
which corresponds to assuming that $\lnMobs$ near the selection threshold follows a normal distribution centered on $\lnM$ with standard deviation $\sigmalnM$.\footnote{Here $\sigmalnM$  is a shorthand $\sigma_{\lnMobs|M}$ and is our definition of scatter throughout this paper.  In other work, sometimes the scatter is defined as the standard deviation of $\lnM$ at a given $\Mobs$, i.e., $\sigma_{\lnM|\Mobs}$, which differs from our definition by the slope of the observable--mass relation.}  In practice, the exact function form of $\phi$ makes little difference as long as it allows for a gradual transition from 0 to 1 (e.g., a ramp function), reflecting the scatter between $\lnMobs$ and $\lnM$.

In this paper, we choose $\Mobsth$ and $\sigmalnM$ as our nuisance parameters.  For a given scatter, we choose an $\Mobsth$ so that the cluster number counts are the same as in the case of zero scatter.  For example, for $\sigmalnM=0.2$, we use $\Mobsth = 1.03 \times 10^{14}~\hiMsun$, which gives the same cluster number counts as $\sigmalnM=0$ and $ M\ge10^{14}~\hiMsun$.  Hereafter we denote such a selection as $\Meqth = 10^{14}~\hiMsun$ (having equal counts as $M\ge10^{14}~\hiMsun$ in the case of no scatter).

For a given survey volume $V$, the expected total number counts are given by
$\avg{N} = \bar{n} V$.  For the convenience of readers, in Appendix~\ref{app:counts} we provide survey volumes and predicted cluster counts for various assumptions about survey properties.

The noise of number counts is contributed by the Poisson noise ($\bar{n}V$) and the halo sample variance \citep{HuKravtsov03}: 
\beq
{\rm Var}[N] = \bar{n}V + V^2 {\bar n}^2 {\bar b}^2 \sigma_V^2  ~,
\label{eq:counts_sv}
\eeq
where $\bar{b}$ is the mean halo bias associated with the cluster sample,
\beq
\bar{b} = 
\frac{1}{\bar{n}}\int dM \frac{dn}{dM} b(M) \phi(\Mobsth | M)  ~,
\eeq
and $\sigma_V$ is the rms density fluctuation in the survey volume,
\beq
\sigma_V^2 = \int 
\frac{k^2 dk}{2\uppi^2} W^2(k,R) P_{\rm lin}(k)  ~.
\eeq
Here $P_{\rm lin}(k)$ is the linear matter power spectrum,
$W(k,R)$ is the window function of the survey, and $R = (3 V / 4\uppi)^{1/3} $.  We use a real-space top-hat window function 
\beq
W(k, R) = \frac{3[\sin(kR)-kR \cos(kR)]}{(kR)^3}   ~.
\eeq
For a large survey volume, $\sigma_V^2 \propto 1/V$.

The observed cluster number counts depend on $\Mobsth$ and $\sigma_{\lnM}$, both of which are degenerate with cosmological parameters.  For example, a larger number count can be achieved by   a higher $\sigma_8$, a lower halo mass at the selection threshold, or a larger scatter.  Therefore, to constrain cosmological parameters, we need external observables to break the degeneracy.  In this work we use stacked weak gravitational lensing, which can be interpreted as calibrating the mean mass of clusters above the $\Mobs$ threshold.  We assume that the scatter $\sigmalnM$ is known with some uncertainty, and we investigate the sensitivity of cosmological constraints to the uncertainty in $\sigmalnM$.  Additional observables that could be used to constrain $\sigmalnM$ include the cluster auto-correlation or cross-correlation with galaxies \citep{Baxter16, Salcedo20}, or comparisons of mass estimates from different observables \citep{Bleem20}.

\subsection{Weak gravitational lensing of galaxy clusters}\label{sec:lensing}

\begin{figure*}
    \centering
    \includegraphics[width=\textwidth]{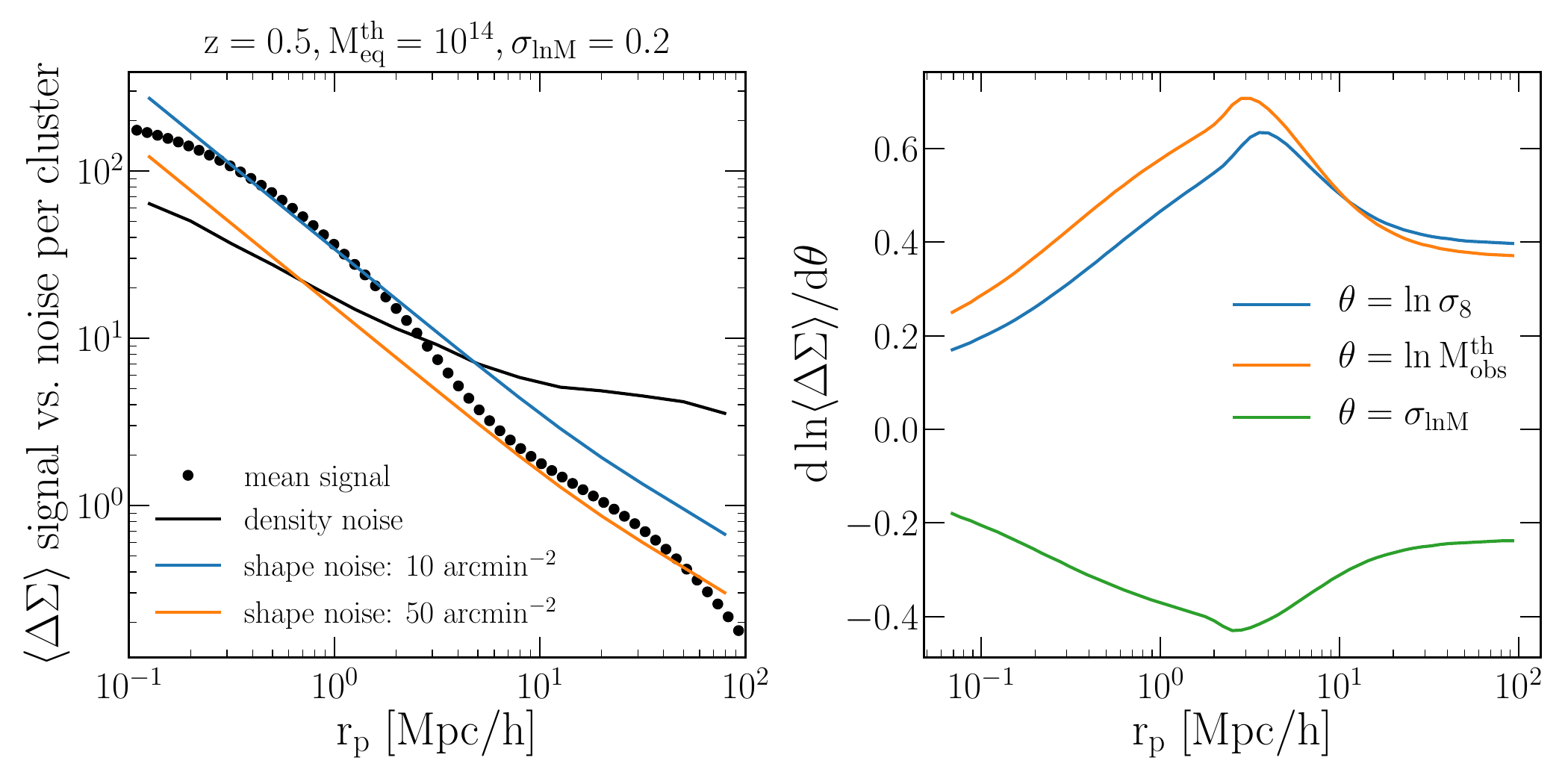}
    \caption{
    {\rm Left}:
    stacked cluster lensing signal (points) and noise per cluster (solid curves), for a selection threshold $10^{14}~\hiMsun$ at $z=0.5$.
    The noise is contributed by density fluctuations (black) and shape noise (blue and orange, which correspond to source densities of 10 and 50 arcmin$^2$ at $z=1.25$).
    In a sample of $N_{\rm cl}$ clusters, each of the noise curves would be lower by $N_{\rm cl}^{-1/2}$.
    {\rm Right}:
    derivatives of the mean lensing signal with respect to $\ln\sigma_8$, $\ln\Mobsth$, and $\sigmalnM$.  The scale dependence of these derivatives is very similar, indicating the strong degeneracy between these parameters in determining the stacked cluster lensing signal.
    }
    \label{fig:DS}
\end{figure*}
\begin{figure*}
    \centering
    \includegraphics[width=\textwidth]{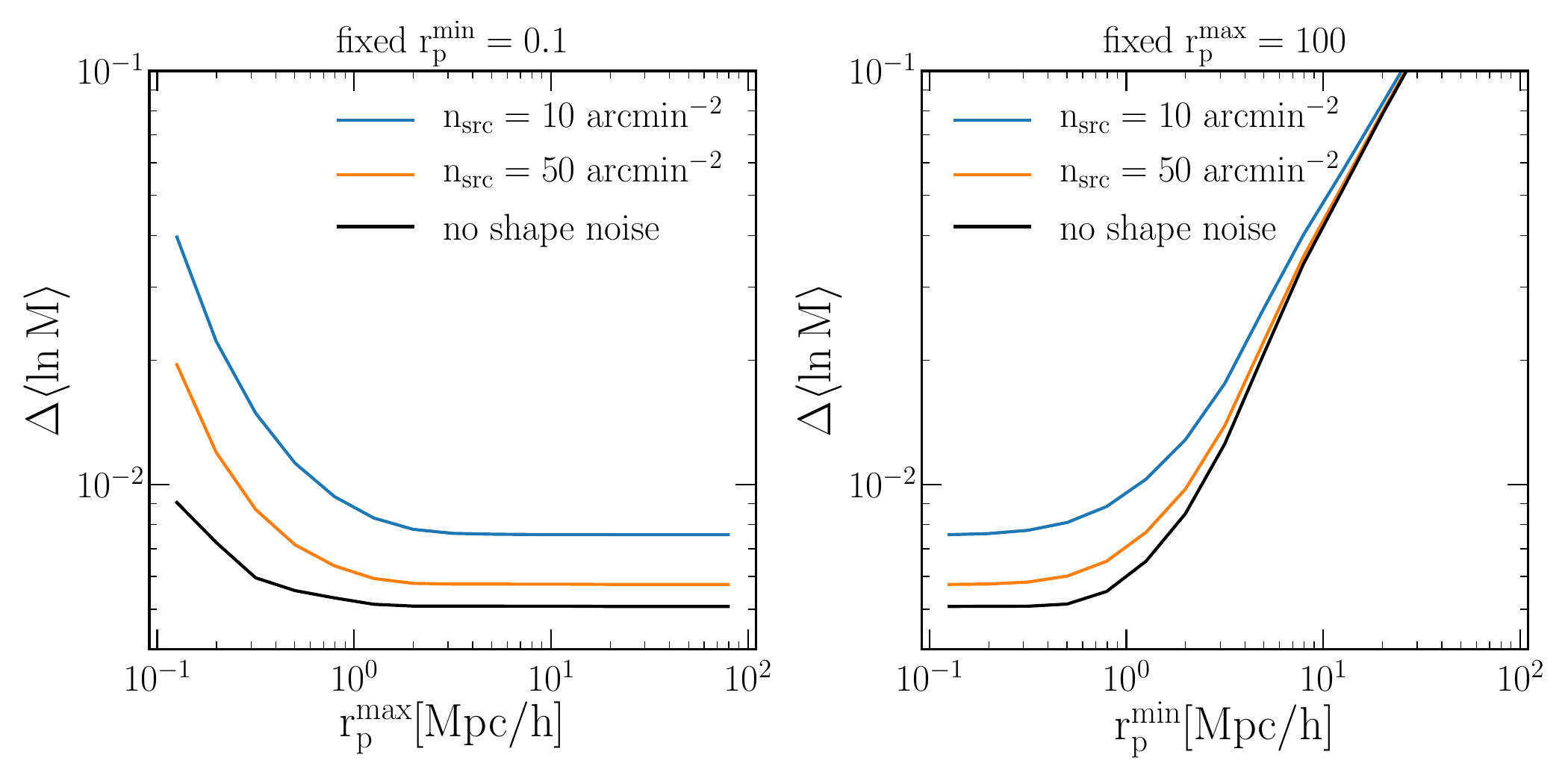}
    \caption{Constraints on the cluster mean mass using stacked lensing, with fixed $\sigma_8$ and scatter.  The left-hand panel corresponds to fixing the smallest scale and progressively including larger scales, while the right-hand panel corresponds to fixing the largest scale and progressively including smaller scales.  The blue and orange curves correspond to source densities of 10 and 50 arcmin$^{-2}$, while the black curve corresponds to an infinite source density (no shape noise). The curves on the left-hand panel drop more rapidly than those on the right-hand panel, indicating that the one-halo regime ($< 3\ \hiMpc$) contains most of the information. The right-hand panel shows that omitting scales below  $\rp = 0.5\ \hiMpc$ would lead to a negligible loss of information.
    }
    \label{fig:Mmean_rp_split}
\end{figure*}

The weak lensing signal generated by stacking a sample of clusters can be expressed in terms of the correlation function between the clusters and the matter density field, $\xicm$:
\beqa
\avg{\DS(\rp)} =& \bar{\rho}\bigg[\frac{4}{\rp^2}\int_0^{\rp} {\rp^\prime}\dd{\rp^\prime} \int_0^\infty  
\dd\chi\ \xicm\left(\sqrt{{\rp^\prime}^2+\chi^2}\right)
\\ 
& - 2 \int_0^\infty \dd\chi\ 
\xicm\left(\sqrt{\rp^2+\chi^2}\right)
\bigg]  \ ,
\label{eq:DS}
\eeqa
where
\beq
\xicm(r) = \int dM  \frac{dn}{dM}  \ 
\xihm(r|M) \ \phi(\Mobsth | M)   ~ ,
\eeq
and $\xihm$ is the correlation function between dark matter and halos of mass $M$.
We approximate $\xihm$ by the maximum of the one-halo and two-halo contribution \citep[e.g.,][]{HayashiWhite08}:
\beq
\xihm = {\rm max}(\xi_{\rm 1h}, \xi_{\rm 2h}) ~,
\eeq
where
\beqa
\xi_{\rm 1h} &= \frac{\rho_{\rm NFW}(M,r)}{\bar{\rho}} - 1  \\
\xi_{\rm 2h} &= b(M) \xi_{\rm lin}(r)  ~, 
\eeqa
and $\rho_{\rm NFW}(r)$ is the halo density profile described by \cite{NFW97}. Here $b(M)$ is the halo bias, and $\xi_{\rm lin}$ is the matter correlation function calculated from the linear matter power spectrum. We adopt the concentration--mass relation from \cite{Correa15}, which is cosmology independent.

We have provided a detailed treatment of cluster lensing covariance matrices in \citet{Wu19}.   To summarize, the covariance matrices for cluster gravitational lensing are contributed by 
\begin{itemize}
    \item Shape noise, which is inversely proportional to the number of source galaxies.
    \item Large-scale structure noise contributed by the line-of-sight structure outside galaxy clusters, which dominates the noise at large scales.
    \item Intrinsic variation of halo density profiles, which contributes to the noise at small scales.
\end{itemize}
For most Stage III optical imaging surveys, the last component is subdominant, but for upcoming surveys like {\rm Roman Space Telescope}'s High Latitude Survey (HLS), this component will be significant.

We analytically calculate the covariance matrices assuming Gaussian random fields (Equation (22) in \citealt{Wu19}), which account for the shape noise and the large-scale structure noise. The small-scale contribution from halos is calculated using the Abacus simulations \citep{Garrison18}.  To validate the combination of analytical Gaussian calculations and the numerical non-Gaussian calculations, we use ray-tracing simulations provided by \citet{Takahashi17}.  We do not use lensing weights in the covariance calculation; in the shape-noise-dominated regime, the optimal weight is given by $\Sigma_{\rm crit}^{-2}$ \citep[see, e.g.][]{Sheldon04, Mandelbaum05, Shirasaki18}, but in our calculation, shape noise does not always dominate, and there is no known analytical optimal weight.

The left-hand panel of Figure~\ref{fig:DS} shows the stacked cluster lensing signal vs.\ noise for our baseline example:  $\zlens=0.5$, $\Meqth = 10^{14}~\hiMsun$, and $\sigmalnM=0.2$.  The points show the signal, and the solid curves show the noise {\em per cluster}.  For a sample of $N_{\rm cl}$ clusters, the noise level scales approximately with $1/\sqrt{N_{\rm cl}}$.   We use five logarithmic radial bins per decade, and we assume that the source galaxies are located at $\zsrc=1.25$.  The black curve shows the contribution of density fluctuations (the sum of large-scale structure and small-scale halo density profiles).  The blue and orange curves show the shape noise for source densities $\nsrc$ of 10 and 50 arcmin$^{-2}$, which approximate the capabilities of Stage III and Stage IV experiments.  This color scheme will be used repeatedly in this paper.

If the shape noise is negligible (black curve), the noise from density fluctuations is still higher than the signal of a single cluster at large scales.  For a Stage-IV level of shape noise (orange), we will have a sufficient signal-to-noise ratio for individual massive clusters at small scales, but for the majority of the clusters and large scales, we will need to stack the signals of multiple clusters.  In Section~\ref{sec:surveys}, we will implement the detailed survey assumptions, which can be calculated by re-scaling the various components in the covariance matrix. For example, the shape noise is proportional to $\rm \avg{\Sigma_{crit}^2}/\nsrc$ (which can be calculated from the source redshift distribution), the contribution of density fluctuation is approximately independent of the source redshift, and the entire covariance matrix is inversely proportional to the survey area.

The right-hand panel of Figure~\ref{fig:DS} shows the derivatives of $\ln\avg{\DS}$ with respect to the model parameters: $\ln\sigma_8$, $\lnMobsth$, and $\sigmalnM$.  The derivative with respect to $\lnMobsth$ is the easiest to understand: increasing $\lnMobsth$ increases the mass scale at the threshold, and thus $\DS$ increases at all scales.  The one-halo regime reflects the change of the density profile with mass and concentration; since $\DS$ is affected by the total mass density inside the aperture, its sensitivity peaks approximately at the outer edge of the one-halo regime.  The sensitivity in the two-halo regime is associated with the mass-dependence of halo bias.

The derivative with respect to $\ln\sigma_8$ is very similar to that with respect to $\lnMobsth$: increasing $\ln\sigma_8$ also increases the mass of all halos.  We choose a concentration--mass relation that is independent of cosmology; if the concentration--mass relation depends on cosmology, the small-scale profile will have larger differences.  On the other hand, increasing the scatter decreases the mass of halos near the selection threshold and thus leads to a lower $\DS$. The similarity between the derivatives indicates that these parameters are degenerate with each other, and thus lensing alone cannot constrain $\sigma_8$.  We will show that with an external prior on scatter, combining lensing and number density can break the degeneracy between $\sigma_8$ and mass scales.

\section{Constraining power per co-moving volume}\label{sec:constraints_per_volume}

\subsection{Constraining cluster mean mass using gravitational lensing}\label{sec:Mmean_constraints}

In cluster lensing analyses, a common strategy is to use gravitational lensing to constrain the mean mass of clusters, and then combine the mean mass and and number density to constrain cosmological and nuisance parameters \citep[e.g.,][]{Costanzi19SDSS,DESY1CL}.  As shown in Figure~\ref{fig:DS}, the derivatives of $\DS$ with respect to threshold, scatter, and $\sigma_8$ have similar shapes; therefore, we cannot constrain these parameters simultaneously.

We define the mean logarithmic mass of a cluster sample as as 
\beq
\avg{\lnM} = \frac{ \int d M \frac{dn}{dM} \phi(\Mobsth | M) \lnM  }
{\int d M \frac{dn}{dM}  \phi(\Mobsth | M) 
}  \ .
\eeq

We first examine a case in which we fix $\sigma_8$ and scatter and constrain the mean mass.  The constraints on the mean mass can be estimated by
\beq
\big( \Delta\avg{\lnM} \big) ^{-2} = 
\frac{\partial \avg{\DS}^{T} }{\partial \avg{\lnM}} 
{\mathcal C}^{-1}
\frac{\partial \avg{\DS} }{\partial \avg{\lnM}} ~,
\eeq
where ${\mathcal C}$ is the lensing covariance matrix, and 
\beq
\frac{\partial \avg{\DS} }{\partial \avg{\lnM}} = \frac{\partial \avg{\DS} }{\partial \lnMobsth} \frac{\partial \lnMobsth }{\partial \avg{\lnM}}  ~.
\eeq

Figure~\ref{fig:Mmean_rp_split} shows $\Delta\avg{\lnM}$ using small scales vs.\ large scales.  We again use our baseline example, $\zlens=0.5$,  $\Meqth=10^{14}~\hiMsun$, $\zsrc=1.25$, survey volume 1 $(\hiGpc)^3$.  The blue and orange curves correspond to source densities of 10 and 50 arcmin$^{-2}$, and the black curves correspond to an infinite source density (no shape noise).  A source density of 50 arcmin$^{-2}$ is very close to the case without shape noise.  In the left-hand panel, we fix the smallest radius to 0.1 $\hiMpc$ and progressively increase the largest scale used, denoted by $\rpmax$.  When $\rpmax$ increases, the constraints improve up to $\sim1\ \hiMpc$, above which including larger-scale information does not add further information.  This indicates that most of the information comes from the one-halo regime.

In the right-hand panel, we fix the largest radius to 100 $\hiMpc$ and progressively include small-scale information down to $\rpmin$.  When $\rpmin$ decreases, the constraints improve quickly, again indicating that small scales contain more information.  Below 0.5 $\hiMpc$, the information saturates, which indicates that discarding scales smaller than $\sim$ 0.5 $\hiMpc$ would lead to little loss of information.  This is encouraging because using scales below 0.5 $\hiMpc$ is more challenging due to  systematic uncertainties such as baryonic effects, cluster mis-centering, and weak lensing boost factors.  By construction, the right end of the left-hand panel matches the left end of the right-hand panel, both corresponding to including information from all scales.  Using scales $\rp \gtrsim 3\ \hiMpc$ yields an error roughly twice that of scales $\rp \lesssim 3\ \hiMpc$; therefore, although the two-halo regime is less constraining, it is competitive enough to allow for a useful check from scales where the underlying physics and observational systematics are quite different from those in the one-halo regime.

In practice, fitting the mean lensing profile to a mean mass can lead to small bias, which can be corrected for using simulations \citep[e.g.,][]{McClintock19}.  In addition, one sometimes fits the mean mass and concentration simultaneously \citep[e.g.,][]{Applegate14}.  We find that we only need a modest prior on concentration to avoid degradation in the mass constraint, and therefore we adopt a fixed concentration--mass relation in this work.

\subsection{Combining cluster number density and lensing to constrain $\sigma_8$}
\label{sec:counts_lensing}

We investigate the constraining power on $\sigma_8$ by combining number counts and gravitational lensing observed at different scales.  We start by assuming a survey volume of 1 $(\hiGpc)^3$, a galaxy cluster sample selected with $\Meqth=10^{14}~\hiMsun$ in a narrow lens redshift bin centered on $\zlens$, and a narrow source redshift bin at $\zsrc = 2.5 \zlens$.

We calculate the Fisher matrix for the data vector ${\bf D} = (N, \avg{\DS})$, 
and model parameters $p$ =  ($\lnsigeight$, $\lnMobsth$, $\sigmalnM$)
\beq
F_{ij} = \frac{\partial {\bf D}}{\partial p_i} {\mathcal C}^{-1}  \frac{\partial {\bf D}}{\partial p_j} \ .
\eeq
To add a prior $\sigma_{\rm prior}$ to parameter $i$, we add $\sigma_{\rm prior}^{-2}$ to the diagonal element $F_{ii}$.
The marginalized constraint on parameter $p_i$ is given by 
\beq
\sigma(p_i) = \sqrt{(F^{-1})_{ii}}  \ .
\eeq
We assume no covariance between number density and lensing.  We have verified this assumption using the Abacus simulations (shown in Appendix~\ref{app:abacus_counts_lensing}).

\begin{figure*}
    \centering
    \includegraphics[width=\textwidth]{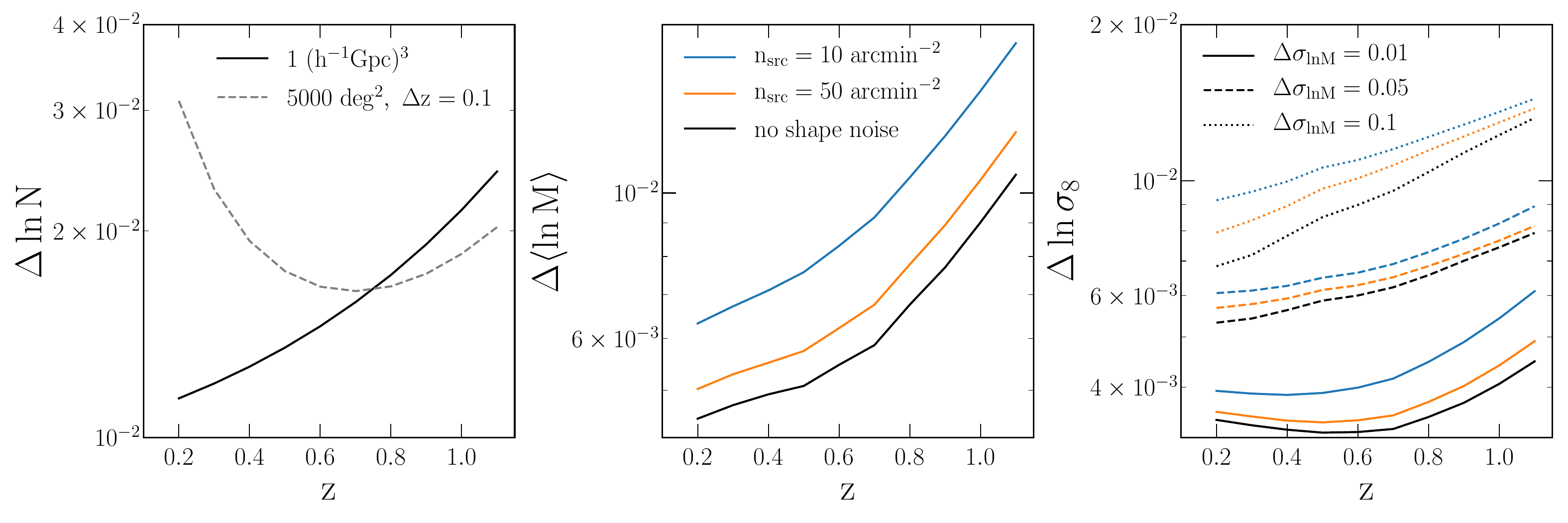}
    \caption{    
    Error bars of $\ln N$ (left-hand),  $\avg{\lnM}$ (middle), and $\lnsigeight$ (right-hand) from a 1 $(\hiGpc)^3$ volume at various redshifts, for clusters selected with $\Meqth = 10^{14}~\hiMsun$.  In the left-hand panel, we show a dashed curve associated with a fixed area  (to be compared with Figure~\ref{fig:constraints_surveys}). In the middle panel, we use the lensing signal between 0.1 and 100 $\hiMpc$ and assume three levels of shape noise. In the right-hand panel, we combine the information from both number counts and lensing, additionally assuming three levels of prior on the scatter. A source density of 50 arcmin$^{-2}$ and a prior of 0.01 scatter is very close to the statistical limit of no shape noise and perfectly known scatter. The constraining power at a fixed volume is slightly better at low redshifts because of the higher cluster co-moving density. 
    }
    \label{fig:constraints_sigma8_z}
\end{figure*}

Figure~\ref{fig:constraints_sigma8_z} demonstrates the constraining power of a 1 $(\hiGpc)^3$ volume survey at various redshifts.  This is not a direct forecast for a sky survey, in which different redshift ranges have different volumes.  The left-hand panel corresponds to the uncertainties in cluster number counts, which are contributed by Poisson noise and the sample variance (Section~\ref{sec:number}) and are determined by the mass and the volume.   The solid curve corresponds to a fixed volume, 1 $(\hiGpc)^3$, which is also used in the other two panels.  To compare with the calculations with fixed area (Figure~\ref{fig:constraints_surveys} below), we add a dashed curve corresponding to 5000 $\deg^2$, $\Delta z = 0.1$, at all redshifts.

The middle panel shows the constraints on the mean mass using the lensing signal from 0.1 to 100 $\hiMpc$, assuming three levels of shape noise (indicated by different colors).  The right-hand panel shows the constraints on $\sigma_8$ combining number density and lensing, for three levels of shape noise (different colors) and three levels of prior on scatter (different line styles).  For a fixed volume, the constraining power on $\sigma_8$ is slightly better at low redshifts due to the higher cluster density and larger source density for lensing.  Nevertheless, the trend is very weak, indicating that the constraining power per volume is approximately constant with redshift.  In all cases, the constraint on $\sigma_8$ for $\Delta\sigmalnM=0.01$ is very close to that of a perfectly known scatter (not shown).

\begin{figure}
    \centering
    \includegraphics[width=0.5\textwidth]{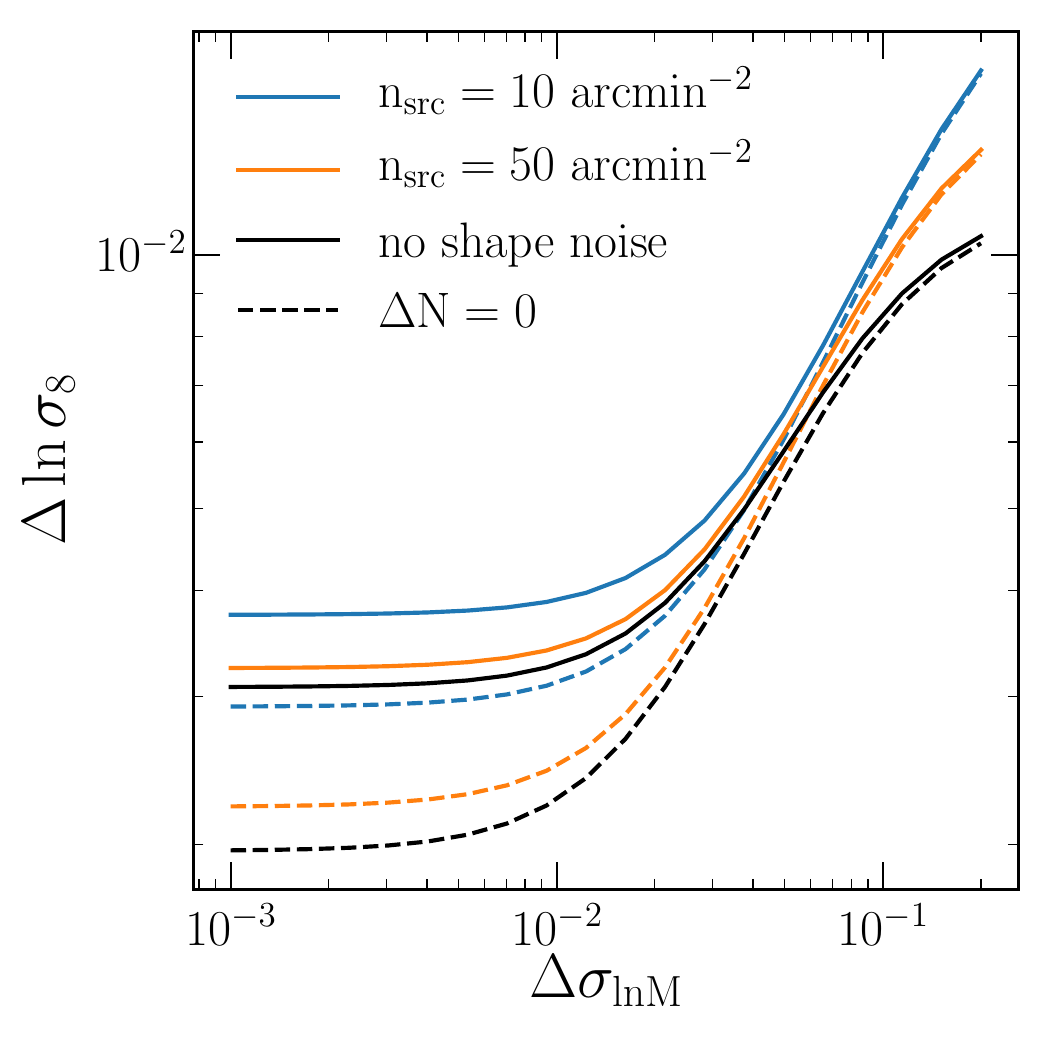}
    \caption{Dependence of $\sigma_8$ constraints on the prior on scatter $\Delta\sigmalnM$, for $z=0.5$, $\Meqth=10^{14}~\hiMsun$, and  $V=1~(\hiGpc)^3$, for three levels of shape noise.  The solid curves correspond to the number count uncertainties associated with a 1 $(\hiGpc)^3$ volume, while the dashed curves correspond to negligible number count uncertainty.  When $\Delta\sigmalnM \approx $ 0.05, improving the prior on scatter plays a more important role than either increasing the cluster sample or increasing the source density.
    }
    \label{fig:constraints_scatter_prior}
\end{figure}

Figure~\ref{fig:constraints_scatter_prior} shows the constraints on $\lnsigeight$ as a function of the prior on scatter ($\Delta\sigmalnM$), for three levels of lensing uncertainties represented by different colors, for our baseline example $z=0.5$, $\Meqth=10^{14}~\hiMsun$, and $V=1\ (\hiGpc)^3$.  The solid curves assume the number count uncertainties associated with the 1 $(\hiGpc)^3$ volume, while the dashed curves assume negligible number count uncertainties ($\Delta N = 0$).   The latter would be relevant if we had a very large volume but only a small subset of clusters had high-quality lensing measurements.  When the scatter is well constrained (left-hand side of the solid curves), increasing the source number density from 10 to 50 arcmin$^{-2}$ produces a modest  (13\%) improvement in the $\sigma_8$ constraint, while eliminating shape noise entirely produces little further gain.  If $\Delta N = 0$ (dashed curves), the effect of shape noise is somewhat larger.  For $\Delta\sigmalnM \lesssim 0.01$, number count and lensing uncertainties dominate the $\sigma_8$ uncertainty.  For $\Delta\sigmalnM \gtrsim 0.02$, the uncertainty in scatter significantly inflates the $\sigma_8$ error, and reducing weak lensing or number count uncertainties has little impact unless $\Delta\sigmalnM$ can also be reduced.

\section{Dependence of parameter constraints on survey conditions}\label{sec:surveys}
\begin{figure}
    \centering
    \includegraphics[width=0.5\textwidth]{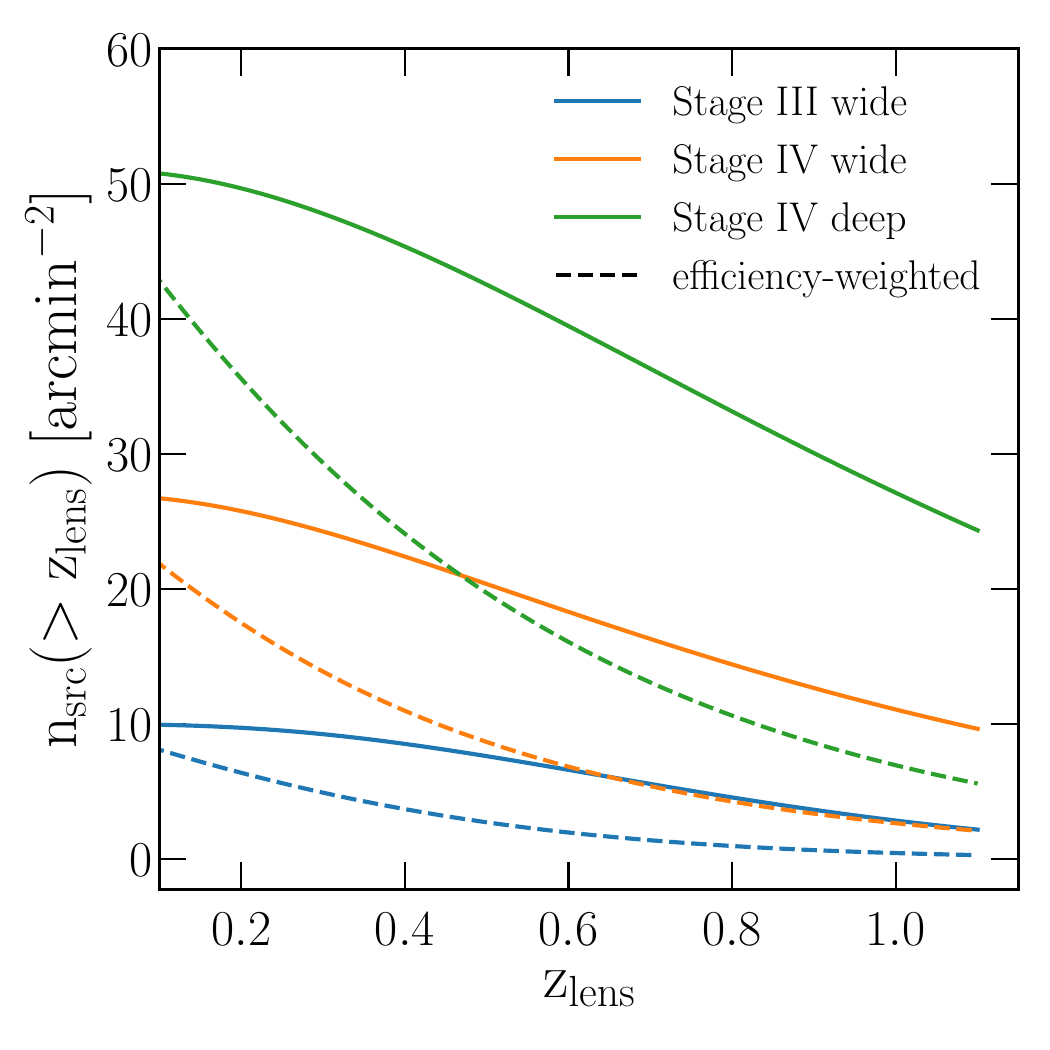}
    \caption{
    Surface number density of source galaxies above a given lens redshift, for various survey assumptions.  The solid curves correspond to the total galaxy number densities above the lens redshift, while the dashed curves are the number densities weighted by the lensing efficiency.
    }
    \label{fig:source_redshift}
\end{figure}
\begin{figure*}
    \centering
    \includegraphics[width=\textwidth]{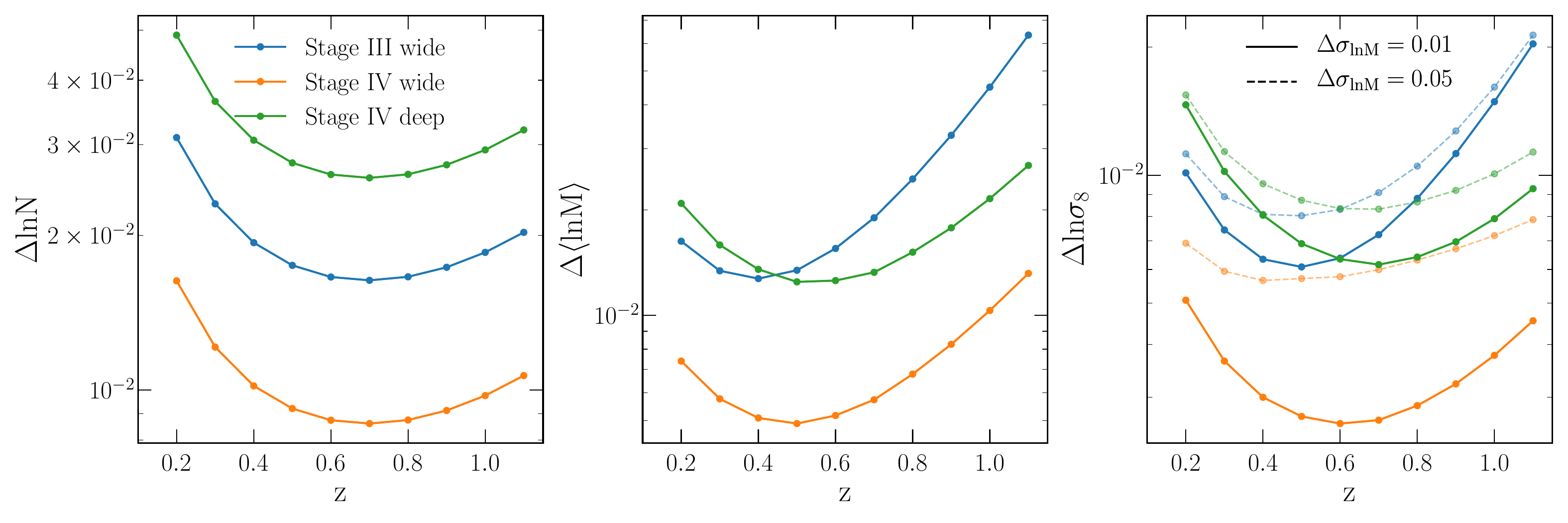}
    \caption{Constraining power for three assumptions about survey properties (see the text), {\rm per redshift bin} of $\Delta z = 0.1$.
    {\rm Left}: fractional uncertainties of cluster number counts, which are inversely proportional to the square root of the survey area.
    {\rm Middle}: fractional uncertainties of cluster mean mass constrained by weak lensing, which are determined by both survey area and source redshift distribution.
    {\rm Right}: fractional constraints on $\sigma_8(z)$, combining number density and lensing, with priors on $\sigmalnM$ of 0.01 (solid) or 0.05 (dashed).
    }
    \label{fig:constraints_surveys}
\end{figure*}
\begin{figure*}
\centering
\includegraphics[width=\textwidth]{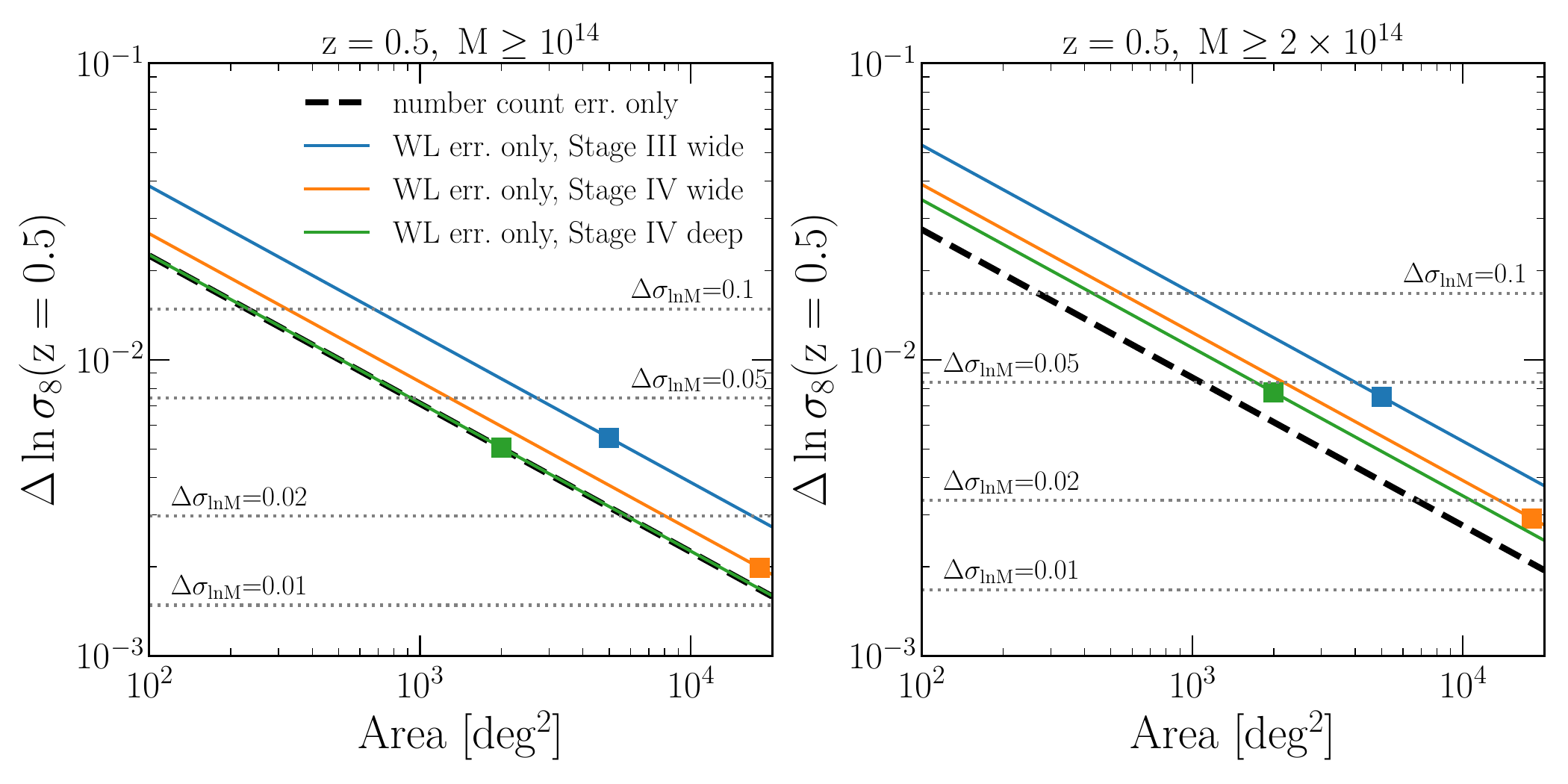}
\includegraphics[width=\textwidth]{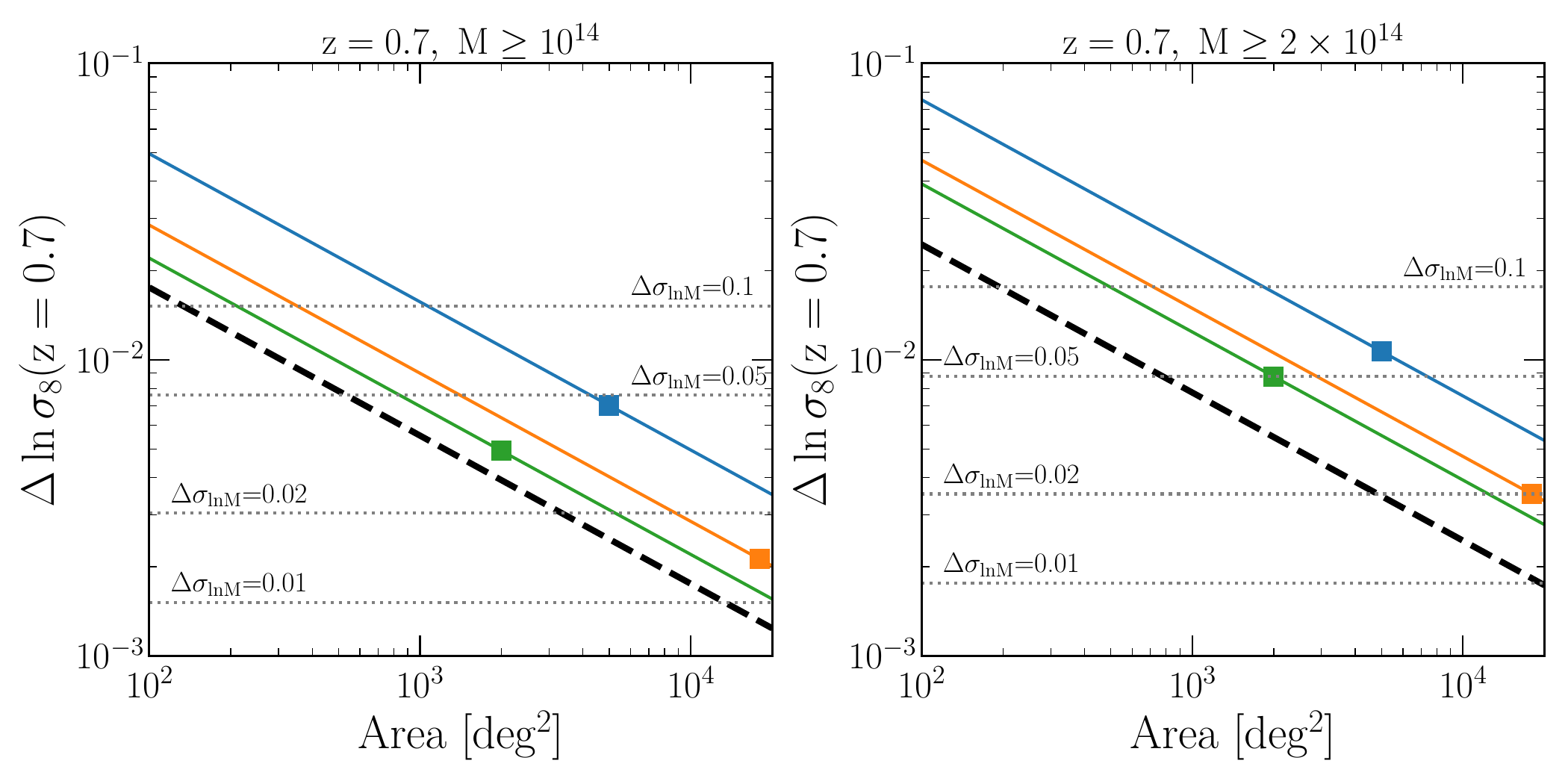}
\caption{Relative importance of the uncertainties related to cluster number counts, lensing lensing, and scatter (external prior).  We show their impacts on $\Delta\lnsigeight$ as a function of survey area, for $z=0.5$ (upper) and $z=0.7$ (lower), and selection thresholds $10^{14}$ (left) and $2\times10^{14}~\hiMsun$ (right).
Here we focus on one source of uncertainty at a time and assume the other two are negligible. The intersection between lines indicates when the impact of the uncertainties becomes comparable.  For example, in the upper left panel for Stage IV deep (green), number count and weak lensing errors contribute almost equally to the $\sigma_8$ uncertainty, and for a $1000$ deg$^2$ survey they have the same impact as $\Delta\sigmalnM = 0.05$. 
In all panels, the blue, green, and orange squares show our assumed survey areas for Stage III wide, Stage IV deep, and Stage IV wide, respectively, and the solid lines show forecasts that assume the same source densities but different survey areas.
}
\label{fig:constraints_surveys_area}
\end{figure*}
\begin{figure}
\includegraphics[width=0.5\textwidth]{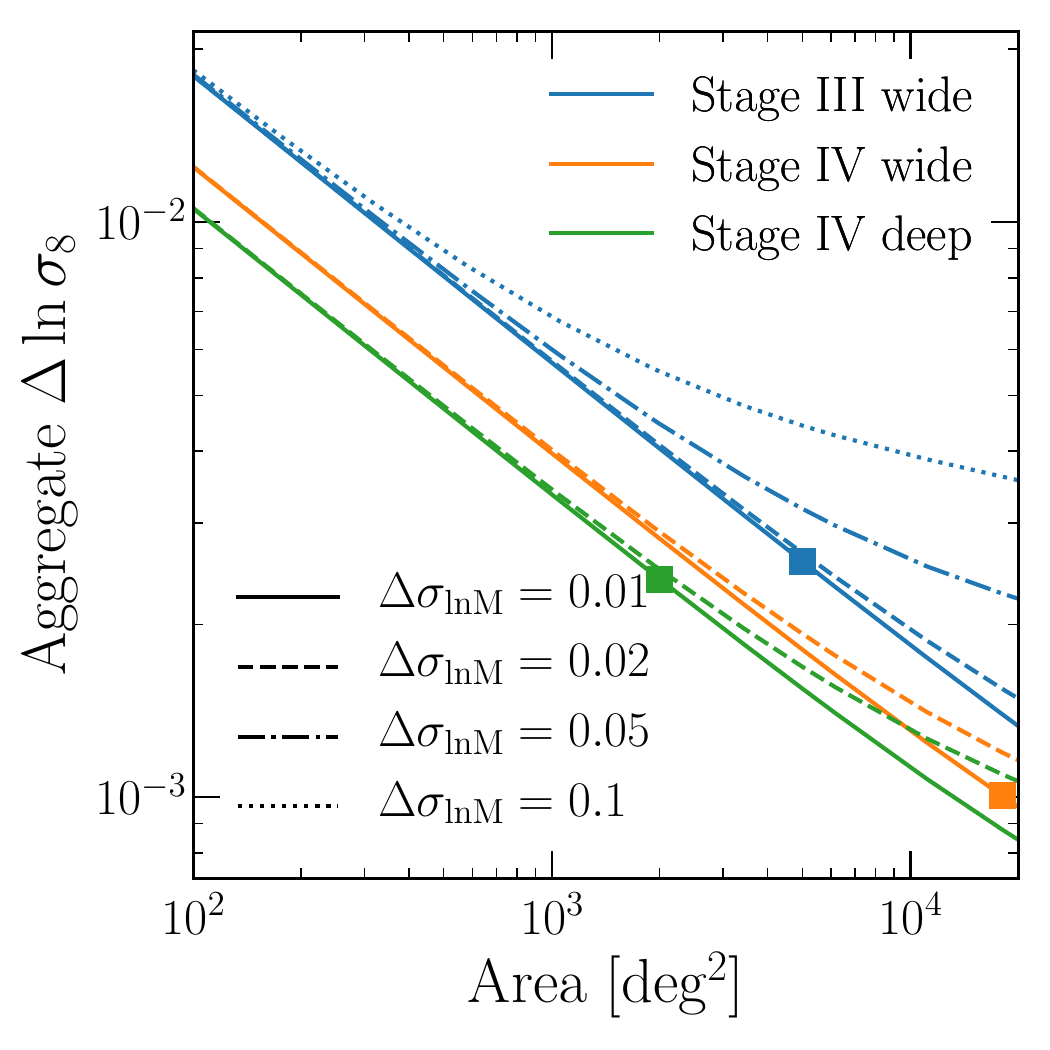}
\caption{Aggregate constraints on $\sigma_8$ combining $0.2\le z \le 1.1$, assuming the information in each redshift bin of $\Delta z=0.1$ is independent.  We assume the source density for various survey assumptions while varying the survey area ($x$-axis); squares mark our assumed survey areas.  Different line styles correspond to different levels of prior on scatter.  A tight prior is required so that the constraining power improves with increasing survey area.}
\label{fig:constraints_surveys_aggregated}
\end{figure}

In this section, we calculate the constraining power of various surveys on $\sigma_8(z)$.  We adopt the following illustrative survey parameters:
\begin{itemize}
\item Stage III wide: 5000 $\deg^2$, $\nsrc$=10 arcmin$^{-2}$
\item Stage IV wide: 18,000 $\deg^2$, $\nsrc$=27 arcmin$^{-2}$
\item Stage IV deep: 2000 $\deg^2$, $\nsrc$=51 arcmin$^{-2}$
\end{itemize}
These assumptions are motivated by the survey designs of Dark Energy Survey (DES), Vera C.~Rubin Observatory Legacy Survey of Space and Time (LSST), and {\rm Roman Space Telescope}'s HLS, respectively.  
For Stage III wide and Stage IV wide, we assume that the source distribution is given by
\beq
\psrc(z) \propto z^2 \exp(-(z/z_0)^\alpha)
\eeq
with $z_0 = 0.11$ and $\alpha=0.68$ for Stage III wide, 
and $z_0 = 0.5$ and $\alpha=1.4$ for Stage IV wide.
For Stage IV deep, we use Figure~2 in \citet{Eifler20WFIRSTLSST}.
We normalize the source distribution such that
\beq
\nsrc = \int_0^{\infty} \psrc(z) dz ~.
\eeq
Figure~\ref{fig:source_redshift} shows the cumulative surface number density of source galaxies above a given lens redshift.  The solid curves correspond to directly integrating $\psrc$ above $\zlens$, while the dashed curves correspond to weighting $\psrc$ by the lensing efficiency,
\beq
q(\chi) = \int_{\chi}^{\infty} d\chi' \psrc(\chi') \frac{\chi' - \chi}{\chi'} \ ,
\eeq
where $\chi$ denotes the co-moving distance \citep[see, e.g.,][for a review]{Kilbinger15}. 
The lensing efficiency characterizes the lensing strength at different lens redshifts for a given source distribution.
Here we normalize $\psrc$ to $\nsrc$ instead of 1.  
We note that this quantity is not the commonly used lensing efficiency, but it provides an easy way to compare the lensing statistical power at a given $\zlens$ from different surveys.
For example, for Stage IV deep (green lines), the source density above  $\zlens=0.5$ is $\approx$ 45 arcmin$^{-2}$ (green solid line), but their lensing signal would be equivalent to  $\approx$ 20 arcmin$^{-2}$ for sources at a single $\zsrc \gg \zlens$ (green dashed line).

We assume that all of these surveys have the same $\Mobsth$ and that the uncertainties in lensing measurements are determined only by the source redshift distribution. Therefore, these assumptions are somewhat optimistic for wide surveys like DES and LSST but potentially achievable with deep surveys like HLS because of the high source galaxy density and better deblending of source galaxies.  Figure~\ref{fig:constraints_surveys} shows the fractional uncertainties on the number counts, the mean mass, and $\sigma_8$, {\rm per redshift bin} of $\Delta z = 0.1$.  The points indicate the middle of the bin.  {\rm We assume an independent set of model parameters in each redshift bin.}  This assumption is observationally realistic; because of the evolution of cluster properties, the same $\Mobs$ threshold may not correspond to the same true mass threshold at different redshifts, and the corresponding mass--observable scatter may not be the same either.   For a specified cosmological model, one can combine $\sigma_8(z)$ estimates to obtain a tighter constraint on $\sigma_8(z=0)$, while the errors in $\sigma_8(z)$ indicate the power of the survey to map out the low redshift growth history.

The left-hand panel shows the error bars on number counts.  Since we assume the same cluster selection for these surveys, the uncertainty of number counts is inversely proportional to the square root of the survey area. The three survey assumptions have an area ratio of 1:~2.5:~9, and thus the fractional count error scales as 3:~1.9:~1.

The middle panel shows the uncertainty of mean mass, which is determined by the lensing covariance matrix.  The elements of the covariance matrix are inversely proportional to the number of clusters multiplied by the sum of density noise and shape noise.  When we compare Stage IV wide vs.\ deep, the larger number of clusters of the wide survey outperform the higher source density of the deep survey.  The right-hand panel shows the resulting constraints on $\lnsigeight$.  We assume two levels of prior on the scatter.  With a tight prior on scatter (solid curves), the constraints closely follow the mass uncertainties, while a weak prior significantly degrades the constraints.

Figure~\ref{fig:constraints_surveys_area} isolates the roles of number count uncertainties (dashed), lensing uncertainties (solid), and uncertainties on scatter (dotted horizontal) for determining the constraints on $\sigma_8$.   The uncertainties in number counts and lensing both decrease with area, while the uncertainty on scatter is assumed to come from external experiments and is thus independent of area.  We show two redshifts, $\zlens=0.5$ (upper) and 0.7 (lower), and two selection thresholds, $\Meqth=10^{14}~\hiMsun$ (left) and $2\times 10^{14}~\hiMsun$ (right).  We consider one source of uncertainty at a time and assume the other two are negligible.   The intersections between lines indicate comparable contributions from these sources of uncertainties.  For example, in the upper left panel ($10^{14}~\hiMsun$), for Stage IV wide, the contributions from number count uncertainties (dashed) and lensing uncertainties (green) are comparable for any survey area.  For a 1000 deg$^2$ survey area,  $\Delta\sigmalnM$ would exceed the other contributions for $\Delta\sigmalnM > 0.05$ and be smaller for $\Delta\sigmalnM < 0.05$.  For a 6000 deg$^2$ survey area, $\Delta\sigmalnM$ must be smaller than 0.02 to avoid degrading the $\sigma_8(z)$ constraints.

For a higher mass threshold, the number of clusters at a fixed survey area is smaller.  The number count and weak lensing uncertainties both increase, but the increase for number counts is smaller because the sample variance contribution is almost unchanged (see Equation~(\ref{eq:counts_sv})).  For lensing uncertainties, at $z=0.5$ the greater depth of Stage IV wide produces significant (64\%) gains over Stage III wide, but the change to Stage IV deep produces little further gain (7\%). At $z=0.7$, the benefit of the greater depth is somewhat larger (30\%).

Figure~\ref{fig:constraints_surveys_aggregated} shows the aggregate constraints on $\sigma_8$, assuming the constraints on $\sigma_8(z)$ for $0.2 \le z \le 1.1$ at each redshift bin of $\Delta z = 0.1$ are independent of each other.  For comparing the amplitude of low redshift clustering to extrapolation from the cosmic microwave background assuming $\Lambda$CDM, this aggregate $\sigma_8$ precision is the relevant measure of survey power.  It also indicates the level of weak lensing systematics control required; roughly speaking, the impact of a 0.5\% change in $\sigma_8$ is comparable to that of a 0.5\% multiplicative shear bias. We use the source redshift distributions of the three survey assumptions and vary the survey area (the $x$-axis).  We show four assumptions for the prior on scatter, again allowing independent values of $\sigmalnM$ in each redshift bin.  When the external prior on scatter at each redshift is well constrained (solid curves), the constraining power on $\sigma_8$ keeps improving with increasing area.  When the scatter is not well constrained, the precision on $\sigma_8$ is limited by the prior on scatter; for example, if $\Delta\sigmalnM = 0.1$, increasing the survey area beyond 2000 deg$^2$ yields limited improvement.

\section{Discussion}\label{sec:discusssions}

\subsection{Statistical uncertainties}

Our results highlight a number of key points. First, realizing the statistical precision achievable with current and future cluster weak lensing surveys requires good knowledge of the scatter $\sigmalnM$ between mass proxies and true mass. For a DES-like (Stage III wide, $5000\deg^2$) or HLS-like (Stage IV deep, $2000\deg^2$) survey, uncertainty at the level of $\Delta\sigmalnM = 0.05$ would degrade the aggregate precision on $\sigma_8$ by $\approx$ 20\% relative to a much tighter (0.01) prior (Figure~\ref{fig:constraints_surveys_aggregated}). For an LSST-like (Stage IV wide, $18,000\deg^2$) survey, a 20\% degradation of $\sigma_8$ precision occurs for $\Delta\sigmalnM = $ 0.016. The weak lensing calibration from these surveys constrains the mean cluster mass at the 1\%--2\% level (DES-like, HLS-like) or the 0.5\%--1\% level (LSST-like) in each of the $\Delta z = 0.1$ bins that contributes substantially to the cosmological constraints (Figure~\ref{fig:constraints_surveys}). An alternative mass calibration technique, e.g., based on cluster X-ray properties, would need to be this accurate or better to achieve the cosmological constraints forecasted here.

One approach to constraining $\sigmalnM$ is to use simulations, which may give a reliable estimate of mass--observable scatter even if they cannot predict the mean relation at the required accuracy. For example, hydrodynamic simulations have shown that gas-based mass proxies have a scatter of 10\%--15\%, depending on the radius and dynamic state \citep[e.g.,][]{Kravtsov06, Nagai06, Stanek10, Wu15, Yu15, Hahn17, LeBrun17}, with gas mass, X-ray temperature, and the combination of the two achieving the smallest scatter. Optical mass tracers tend to have a larger scatter, approximately 30\%--50\% for richness and 20\%--30\% for stellar mass \citep[e.g.,][]{Pillepich18sm, Bradshaw20, Anbajagane20}, depending on the selection of cluster members. 
In addition, simulations can also be used to characterize the correlated scatter among different mass proxies, which could be related to projection effects \citep[e.g.,][]{Angulo12, Shirasaki16}.

Another approach is to compare one observable (e.g., optical galaxy richness) to another (e.g., X-ray temperature) that is expected to have substantially smaller scatter \citep[e.g.,][]{RozoRykoff14}. The scatter of the first observable can be estimated by subtracting in quadrature the estimated scatter of the more precise observable, which can in turn be inferred from simulations. Applying this approach to X-ray observations of $\sim 200$ optical clusters from DES, \cite{Farahi19} infer $\sigma_{\rm lnM | \lambda} = 0.30 \pm 0.04_{(\rm stat)} \pm 0.09_{(\rm sys)}$ for clusters selected by richness $\lambda$ from the redMaPPer algorithm. An uncertainty $\Delta\sigmalnM = 0.1$ degrades the achievable DES $\sigma_8$ precision by a factor of 1.4 relative to $\Delta\sigmalnM=0.05$, and therefore improved determination of $\sigmalnM$ is needed to realize the potential of DES measurements. Even with $\Delta\sigmalnM = 0.1$, our forecast of the achievable precision on $\sigma_8$ from the final DES cluster weak lensing data set is $\approx 0.36\%$, a large improvement over the current $\sim 3\%$ precision from DES Year 1 galaxy clustering and weak lensing probes \citep{DESY1KP}. If the effective mass threshold for clusters in the cosmological analysis is $2\times 10^{14}~\hiMsun$ instead of $1\times 10^{14}~\hiMsun$ then the forecast $\sigma_8$ precision is 0.58\% for $\Delta\sigmalnM=0.1$ and 0.43\% for $\Delta\sigmalnM=0.05$. Looking to the future, data from the {\rm eRosita} survey and continuing targeted studies with {\rm Chandra} and {\rm XMM} will be valuable for sharpening knowledge of $\sigmalnM$, and SZ data from the Simons Observatory and eventually CMB-S4 will be valuable for higher-mass clusters.  With large multiwavelength samples, we will be able to perform various cluster selections (e.g., selecting X-ray clusters and measuring weak lensing signals) and cross validations with optical selection.  These analyses could improve our understanding of systematics associated with cluster selections.

In \cite{Salcedo20}, we propose an alternative approach to constraining $\sigmalnM$ based on a combination of the projected cluster--galaxy cross-correlation function $w_{\rm cg}(r)$ and the projected galaxy auto-correlation function $w_{\rm gg}(r)$. For a DES-like survey with a cluster mass threshold of $1\times 10^{14}~\hiMsun$, we forecast an error of $\Delta\sigmalnM \approx 0.1$ in a redshift bin $z = 0.35-0.55$. In combination with weak lensing $\Delta\Sigma(r)$ measurement, we forecast a $\sigma_8$ precision of 0.95\% from this redshift bin and 0.81\% when adding a $z = 0.15-0.35$ redshift bin. These can be compared to this paper's  forecast of 0.26\% for the case of perfectly known $\sigmalnM$ and all redshift bins. Compared to this paper's number count plus $\Delta\Sigma(r)$ approach, the $w_{\rm cg} + w_{\rm gg} + \Delta\Sigma$ analysis proposed in \cite{Salcedo20} is much less sensitive to errors in the cluster space density, and perhaps to the projection systematics discussed below, so it provides a useful complement even if $\sigmalnM$ can be adequately constrained by other means.  We also note that $w_{\rm cg}$ and $\DS$ are both affected by the projection effect, and the combination of these three correlation functions could be an effective way for self-calibrating the projection effect.  We will explore this in future work.

A second point highlighted by our results is that for source densities higher than the Stage III value $\nsrc \sim 10\ {\rm arcmin}^{-2}$ the gains in $\sigma_8(z)$ precision begin to grow more slowly than $\sqrt{\nsrc}$. This is true even if one considers weak lensing errors alone, and at Stage IV depth, the number count uncertainties become comparable to weak lensing uncertainties (Figure~\ref{fig:constraints_surveys_aggregated}). For these reasons, the larger area of a Stage IV wide survey wins over the greater depth of a Stage IV deep survey, if both surveys map clusters down to the same mass threshold and are statistics limited.

This brings us to a third, critical point: the achievable precision for statistics-limited cluster weak lensing surveys is high. Our forecast of  aggregate $\sigma_8$ precision with $\Delta\sigmalnM=0.01$ is 0.10\% for Stage IV wide, 0.24\% for Stage IV deep, and 0.26\% for Stage III wide. \cite{OguriTakada11} and \cite{Weinberg13} conclude that cosmological constraints from cluster weak lensing are competitive with those from cosmic shear using the {\em same} weak lensing data set. We do not re-examine this point here, but we note that Spergel et al.\ (\citeyear{Spergel15}, \S 2.2.3.2, forecast by C.\ Hirata) project 0.12\% precision for $\sigma_8$ from cosmic shear with the {\rm WFIRST} (now {\rm Roman}) HLS, treating $\sigma_8$ as the single free cosmological parameter, as we do here. The 0.24\% precision that we forecast for cluster weak lensing is weaker than cosmic shear; however, because these two probes have different sensitivities to other cosmological parameters and to modified gravity scenarios, the combination should be significantly more powerful than either alone.

\subsection{Systematic uncertainties} \label{sec:systematics}

Realizing this high precision requires excellent control of observational systematics in addition to tight constraints on $\sigmalnM$. For example, multiplicative shear biases from shear calibration uncertainty or photo-$z$ uncertainties must be controlled at about the same fractional level as the aggregate $\sigma_8$ uncertainty to avoid dominating the cosmological parameter error. These effects can be estimated by direct calibration methods. A complementary approach is to describe them by nuisance parameters and constrain them using multiple observables. This strategy may strengthen significantly by combining cluster weak lensing with galaxy--galaxy lensing and cosmic shear, especially when using multiple redshift bins and measurements that traverse the linear and nonlinear clustering regimes. We will explore this topic in future work.

There are several systematic effects that are specific to cluster weak lensing. One of these is cluster mis-centering. The central galaxy is often taken as the center ($\rp=0$) for computing tangential shear profiles. If some central galaxies are misidentified or do not reside at the center of the cluster's dark matter halo, then shear profiles will be shallower than model predictions that ignore mis-centering. \cite{Zhang19} quantified the distribution of mis-centering of DES clusters by comparing peaks of X-ray emission with centers determined by the redMaPPer cluster finder. This distribution is in turn incorporated into model predictions when fitting cosmological models to the DES cluster data \citep{McClintock19,DESY1CL}. Our use of a wide $\rp$ range reduces the impact of mis-centering, which only affects the signal on small scales, and it allows consistency checks between the large-scale and small-scale regimes. Nonetheless, controlling mis-centering systematics at the level of our forecast statistical precision will require further multiwavelength cluster studies and simulation studies to accurately quantify mis-centering effects. Hydrodynamic, radiative cooling, and feedback effects alter the distribution of baryonic matter in clusters relative to dark matter \citep{Henson17}. Uncertainties in these effects introduce a theoretical systematic in predicting weak lensing profiles, with an impact that qualitatively resembles that of mis-centering. Continued investigation using hydrodynamic cosmological simulations with varied feedback prescriptions, and development of parameterized models that can be tested against SZ and X-ray data (e.g., \citealt{Mead20}) will be essential to mitigating this systematic in future cluster cosmology analyses.

Another cluster-specific observational systematic is the dilution of the weak lensing signal by apparent sources that are in fact associated with the cluster itself. Put differently, the true redshift distribution of galaxies in a given photo-$z$ bin will be distorted by the presence of a cluster at the center of the field. The correction for this dilution is often called the ``boost factor.'' For the DES Year 1 analysis, the uncertainty of the boost factor is approximately 2\% \citep{Varga19}. Stage IV experiments should have better photometric redshifts and thus more accurate boost factors \citep{Hemmati19}. Boost factor systematics can be checked by investigating radial trends of the weak lensing relative to model expectations and by constructing source galaxy samples with colors that firmly exclude the cluster redshift.

Incompleteness and contamination of cluster catalogs add non-Gaussian tails to the distribution of $P(\ln\Mobs|\ln M)$. Accurate modeling of these tails using simulations and multiwavelength selection comparisons is necessary for making high-accuracy predictions of number counts and weak lensing signals \citep{Aguena18}.

The single most challenging systematic for cluster weak lensing cosmology is {\em anisotropic} selection bias, which arises because triaxial halos oriented along the line of sight or halos projected against correlated large-scale structure can be incorrectly boosted above the $\Mobs$ threshold. These clusters also have stacked weak lensing signals that are higher than expected for random halos of the same mass, an effect that extends to large scales in the two-halo regime (\citealt{Dietrich14,Osato18,Sunayama20}; H.-Y.~Wu et al.~2021, in preparation). Simulations and observations imply that the effect on weak lensing signals can be as large as 20\%--30\% on some scales (see the above references and \citealt{Herbonnet19,DESY1CL}). Correcting for this bias requires simulations that accurately model the entire cluster selection process, with realistic galaxy populations embedded in realistic large scale structure. Creating and analyzing these simulations is a challenging task, and uncertainties in the correction for anisotropic selection bias could well be the long-term limiting factor in cluster cosmology.

The impact of this systematic can be reduced by using a cluster selection observable that is less affected by projection, such as X-ray luminosity or temperature. For this reason, weak lensing of {\rm eRosita} selected clusters may yield the most powerful Stage IV cosmology results even if the effective mass threshold is higher than that achievable with optical cluster selection. SZ selection may also be a valuable tool, though it can also be affected by anisotropy to some degree. In optical surveys, mass proxies weighted toward central galaxies may be less sensitive to projection than the richness estimators in current use, an approach worth further investigation with simulated catalogs.

Systematics associated with cluster selection are somewhat mitigated by our use of a single $\Mobs$ threshold instead of multiple bins, since clusters with $\Mobs$ far above the threshold will remain in the sample regardless of fluctuations from noise, projection, or anisotropy. However, a large fraction of clusters are close to the threshold, so these effects still matter. In Appendix~\ref{app:bins} we examine the impact of splitting a mass threshold sample into two bins, with the addition of nuisance parameters describing $\Mobs$ and $\sigmalnM$ at the bin boundary. We find that there is no gain in cosmological parameter precision. Examination of multiple thresholds or multiple bins may be valuable for identifying and diagnosing systematics --- at a minimum, cosmological parameters derived from samples with different $\Mobs$ thresholds should be mutually consistent. We are not convinced, however, that there is significant cosmological information in the use of multiple $\Mobs$ bins, unless one adopts an unjustifiably restrictive form for the mass--observable relation such as a power law with constant scatter.

\subsection{Ground vs.\ Space}\label{sec:ground_vs_space}

Space-based observations from {\rm Roman} or {\rm Euclid} offer some significant advantages relative to ground-based imaging such as LSST. The sharp and stable point-spread function available from a space platform allows for tighter control of shear calibration uncertainties and better deblending of sources, an effect that may be particularly important in cluster environments. The combination of space-based IR and ground-based optical photometry allows for better photo-$z$ determinations than either alone. The baseline design of the {\rm Roman} HLS focuses on weak lensing systematics control as the primary requirement, choosing depth, multiband coverage, and multiple dithers and roll angles over the greater statistical power that would come from a wider and shallower survey. The most effective cosmological studies may well come from analyses that use overlap of the HLS with LSST and {\rm Euclid} to calibrate systematic effects, and then leverage the statistics of the larger area surveys. \cite{Eifler20WFIRSTLSST} discuss survey strategies that would enable {\rm Roman} to cover the entire LSST footprint with either a 5 month shallow $W$-band survey that would support deblending and source modeling or a 1.5 yr deep $H$-band survey that would enable independent weak lensing shape measurements.

In this spirit, we note that medium-scale guest observer programs with {\rm Roman} could complement the HLS and substantially amplify its power for cluster weak lensing cosmology. The key factors that motivate such a program are (1) that a single {\rm Roman} field of view is large enough to encompass several megaparsecs around a moderate redshift cluster, and (2) that the typical angular separation of clusters is much larger than this field of view. In the concept described by \cite{Akeson19}, the HLS observations consist of eight or nine exposures of 140 s in each band (Y, J, H, F184).  The characteristic exposure depth used for forecasts (including ours) is taken to be $5\times 140 = 700$ s, as individual points on the sky fall into chip gaps on some exposures. An individual cluster can therefore be observed to the HLS depth in the $H$-band in approximately 1000 s, depending on overheads, dither patterns, and other details. These observations would cover a roughly 0.25 deg radius around each cluster, corresponding to 3.1, 5.8, and 7.7 co-moving $\hiMpc$ at $z = 0.25$, 0.5, and 0.7, respectively. Because the cosmological analysis uses the stacked weak lensing signal, homogeneous coverage around each individual cluster is not necessary.

An observing program totaling 2.6 Ms (1 month) could obtain weak lensing observations of 2600 clusters. These could, for example, be the {\rm eRosita} selected clusters above an $\Mobs$ threshold corresponding to $2\times 10^{14}~\hiMsun$ in the redshift range $z=0.65-0.75$, over $14,000\deg^2$ (see Figure~\ref{fig:counts} below). If we add these clusters to our $2000\deg^2$ Stage IV deep forecast, then the constraint on $\sigma_8(z=0.7)$ improves from 0.60\% to 0.33\%, assuming a mass threshold of $1\times 10^{14}~\hiMsun$ in the HLS region and that $\Delta\sigmalnM=0.01$ for both samples. This would be a substantial gain for a program requiring only about 1/15 of the HLS observing time. The impact on the aggregate $\sigma_8$ precision is small (0.24\% vs.\ 0.22\%) because the HLS observes $\Mobs \geq 1\times 10^{14}~\hiMsun$ clusters at {\em all} redshifts (23,000 total). However, a virtue of cluster weak lensing relative to cosmic shear is the ability to target a specific redshift range (or halo mass range) that could be highlighted for investigation by future data. If tensions with cosmological models appear to peak at low redshift, then the most effective use of a targeted 1 month survey might be to observe all 2600 $\Mobs \geq 4\times 10^{14}~\hiMsun$ {\rm eRosita} clusters with $z = 0.25-0.45$ over $20,000\deg^2$.  Such a survey would sharpen the precision on $\sigma_8(z=0.25-0.45)$ from HLS clusters from 0.63\% to 0.35\%.  A longer-duration cluster survey could achieve correspondingly larger gains.

\begin{figure}
    \centering
    \includegraphics[width=0.5\textwidth]{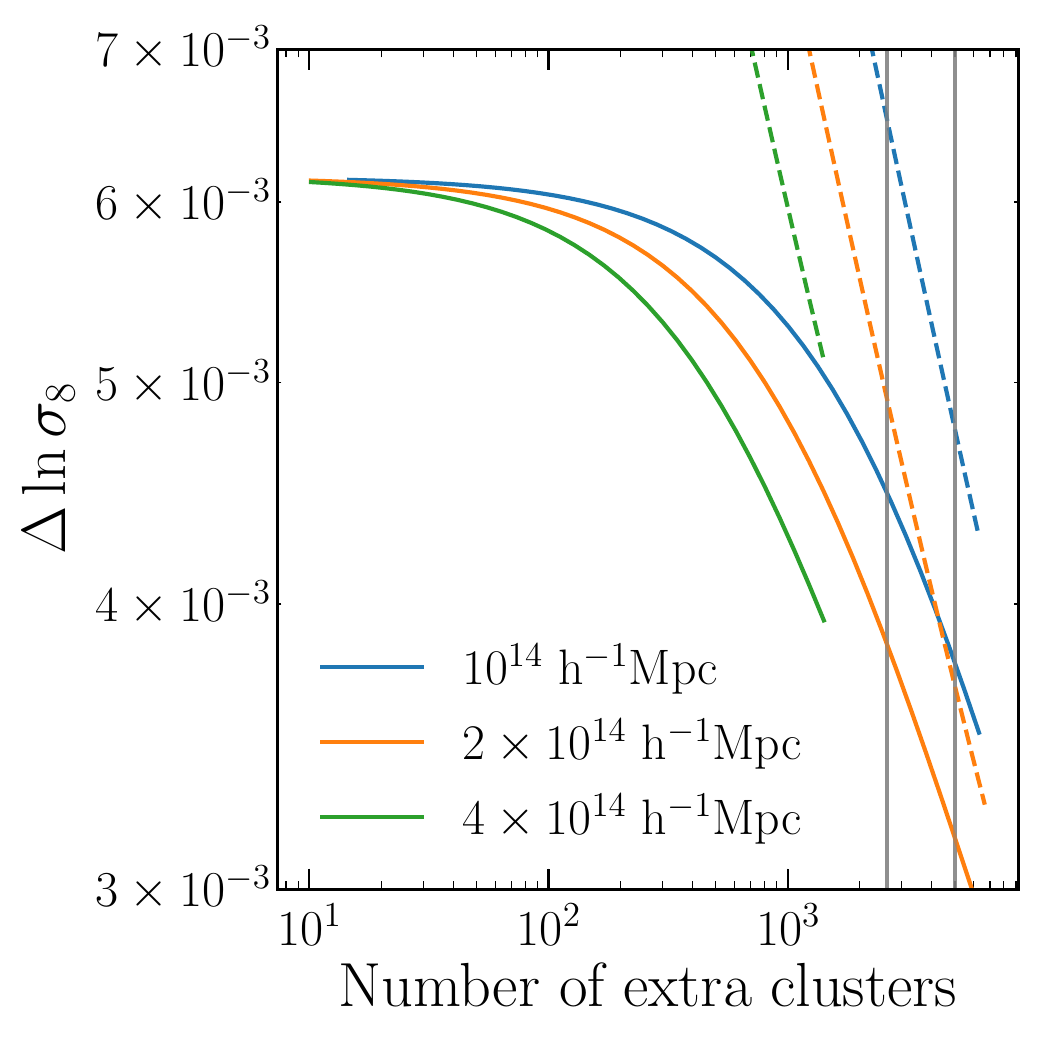}
    \caption{Improvement of $\sigma_8(z=0.7)$ constraints by adding an extended targeted cluster survey using {\rm Roman} on top of HLS, assuming $\Delta\sigmalnM$ = 0.01 for both HLS and targeted observations.  Targeting 2600 clusters above $10^{14}~\hiMsun$ (approximately 2.6 Ms) can improve the $\sigma_8$ constraints by approximately a factor of two at this redshift. Dashed lines show the constraints from the target observations alone.  Vertical lines mark 2600 and 5000 clusters. 
    }
    \label{fig:extended}
\end{figure}

Figure~\ref{fig:extended} shows the improved $\sigma_8$ constraints vs.\ the number of extra clusters observed at $z=0.7$. We show three different $\Mobs$ thresholds corresponding to $\Meqth=10^{14}$, $2\times10^{14}$, and $4\times10^{14}~\hiMsun$. The $4\times 10^{14}$ selection has approximately 1500 clusters in the entire sky in $0.65 < z < 0.75$. If we target clusters individually, a higher mass threshold would lead to a larger improvement in the $\sigma_8$ constraint. The vertical lines indicate 2600 and 5000 clusters. We also added a set of dashed lines that show the constraints achievable with targeted observations alone, before consideration of the HLS.

While even Stage III surveys have the imaging depth needed to identify optical clusters with $\Mobs \approx 10^{14}~\hiMsun$ (see \citealt{Weinberg13}, Figure~27), future studies of systematic effects in cluster selection may show that the threshold for cluster populations that can be effectively used for precision cosmology is higher. In this case, the expected $\sigma_8(z)$ precision from the {\rm Roman} HLS is lower (e.g., Figure~\ref{fig:constraints_surveys_area}), but the relative gains from a targeted survey are larger.

\section{Summary}\label{sec:summary}

Current and upcoming optical imaging surveys will provide unprecedented precision for weak gravitational lensing, making the combination of cluster weak lensing and cluster abundances one of the most powerful probes of the growth of structure. In this work, we explore systematically how the cosmological constraining power of a cluster weak lensing survey depends on the cluster sample size, the lensing source density, and the prior on the scatter between cluster observables and mass. In contrast to most treatments of cluster cosmology, we advocate for an approach, analogous to that of galaxy--galaxy lensing, in which the observables are the mean cluster space density $\bar{n}$ and the mean weak lensing profile $\DS(\rp)$ of clusters above a threshold in an observable mass proxy $\Mobs$, which could be galaxy richness, X-ray temperature, SZ decrement, or some other property. In this formulation, the primary astrophysical nuisance parameter is the scatter $\sigmalnM$ between $\lnMobs$ and the true halo mass $\lnM$ near the threshold. This scatter may vary with redshift, but focusing on a single thresholded sample in each redshift bin avoids the need to adopt a parameterized form of the mean $\Mobs-M$ relation or the dependence of scatter on $\Mobs$. In Appendix~\ref{app:bins} we argue that any statistical gain from considering multiple mass bins or thresholds is minimal at best.

To characterize the constraining power of cluster weak lensing surveys, we focus on the precision in measuring the matter fluctuation amplitude $\sigma_8(z)$ with other cosmological parameters held fixed. In our approach one goes directly from the observed $\bar{n}$ and $\Delta\Sigma(\rp)$ and a prior on $\sigmalnM$ to cosmological parameters such as $\sigma_8$. The mass corresponding to the threshold $\Mobs$ and the mean mass of clusters above the threshold are useful byproducts of the analysis. To build intuition, we first examine expected constraints from a fixed co-moving survey volume of $(1\hiGpc)^3$ at various redshifts, for different assumptions about weak lensing shape noise and the prior on $\sigmalnM$ (Figures~\ref{fig:constraints_sigma8_z}, \ref{fig:constraints_scatter_prior}). We then forecast statistical constraints (Figures~\ref{fig:constraints_surveys}-\ref{fig:constraints_surveys_aggregated}) for three different survey scenarios that correspond approximately to DES (Stage III wide), the {\rm Roman} HLS (Stage IV deep), and LSST (Stage IV wide). Appendix~\ref{app:analytic} provides analytic results that approximately reproduce our numerical results and allow for rapid back-of-the-envelope estimates for alternative survey assumptions or even different methods of estimating cluster mass.

If the scatter is known to $\Delta\sigmalnM = 0.01$, then a survey of $M \geq 10^{14}~\hiMsun$ clusters yields an aggregate precision on $\sigma_8$ of 0.26\%, 0.24\%, and 0.10\% for DES-like, HLS-like, and LSST-like survey parameters, respectively. For a single $\Delta z = 0.1$ redshift bin centered at $z=0.7$, the corresponding errors on $\sigma_8(z=0.7)$ are 0.91\%, 0.83\%, and 0.60\%. The errors on the mean cluster mass in this redshift bin are 1.9\%, 1.3\%, and 0.57\%. If the cluster mass threshold is $2\times 10^{14}~\hiMsun$ instead of $1 \times 10^{14}~\hiMsun$, then the constraints on $\sigma_8(z)$ degrade by roughly a factor of two.

For a statistics-limited survey with the same cluster mass threshold, the greater area of an LSST-like survey (or a {\rm Euclid}-like survey) wins easily over the greater depth of an HLS-like survey. (See \citealt{Eifler20WFIRSTLSST} for a similar comparison of cosmic shear and a detailed discussion of {\rm Roman}+LSST synergies and of possible alternative strategies for the HLS.) However, realizing this advantage requires accurate external knowledge of $\sigmalnM$. For a DES-like or HLS-like survey, a prior with $\Delta\sigmalnM = 0.05$ is sufficient for limited (20\%) degradation of the aggregate $\sigma_8$ constraint, but for an LSST-like survey, the corresponding requirement is $\Delta\sigmalnM = 0.016$. We emphasize that our forecasts allow for a different, independent value of $\sigmalnM$ in each redshift bin and do not assume any specific form of redshift evolution.

Cluster weak lensing surveys can yield stringent tests of cosmological model predictions of matter clustering, and thus of dark energy and modified gravity theories. Our results highlight the importance of constraining $\sigmalnM$, which will require a careful combination of multiwavelength data sets and numerical simulations. For a DES-like survey, \cite{Salcedo20} forecast that the combination of cluster--galaxy cross-correlation and galaxy auto-correlation can achieve $\Delta\sigmalnM \approx 0.1$, which is already sufficient for a 0.5\% constraint on $\sigma_8$ (see Figure~\ref{fig:constraints_surveys_aggregated}).

Realizing the potential of cluster cosmology will require strict control of a variety of observational and theoretical systematics (Section \ref{sec:systematics}), of which anisotropic cluster selection is perhaps the most challenging. Baryonic physics impacts on cluster mass profiles are an important theoretical uncertainty, but limiting the weak lensing analysis to $\rp > 0.5\ \hiMpc$ or even $1.0\ \hiMpc$ produces little degradation of statistical precision (Figure~\ref{fig:Mmean_rp_split}), which suggests that the effects of baryons and cluster mis-centering can be adequately mitigated by including $\Delta\Sigma(\rp)$ measurements out to large $\rp$. Combinations of space-based and ground-based data will be valuable for controlling other observational systematics while leveraging the large area of ground-based surveys. We also note (Section \ref{sec:ground_vs_space}) that relatively short cluster observing programs with {\rm Roman Space Telescope} can achieve tight constraints on $\sigma_8$ in a targeted redshift range, which could be valuable for validating other results or for developing a deeper understanding of departures from $\Lambda$CDM predictions. For example, a roughly 1 month observing program could sharpen the HLS constraint on $\sigma_8(z=0.7)$ by about a factor of two, or make a 0.5\% measurement of $\sigma_8(z=0.7)$ on its own.

If statistical limits can be achieved, then the expected constraints on cosmic structure growth from cluster weak lensing are competitive with and complementary to those from cosmic shear with the same weak lensing data \citep{OguriTakada11,Weinberg13}. The value of cluster weak lensing as an independent probe becomes much higher still in models with modified gravity or decaying dark matter, which predict distinctive signatures in matter clustering as a function of environment, scale, and redshift. If we live in such a universe, then cluster weak lensing will be a critical tool for characterizing its rich phenomenology and revealing its underlying physics.

\section*{Acknowledgments}

We thank the anonymous reviewer for helpful suggestions, Lehman Garrison for providing the Abacus Cosmos simulation suite, and Ying Zu and Chun-Hao To for helpful discussions and comments. 
This work was supported in part by NSF grants AST-1516997 and AST-2009735 and NASA grant 15-WFIRST15-0008.
D.H.W.~acknowledges additional support at the IAS from the W.~M.~Keck Foundation.
The computations in this paper were run on the CCAPP condo of the Pitzer Cluster at the \cite{OSC}.

\appendix

\section{Survey volume and number of galaxy clusters}\label{app:counts}

\begin{figure}
    \centering
    \includegraphics[width=0.5\textwidth]{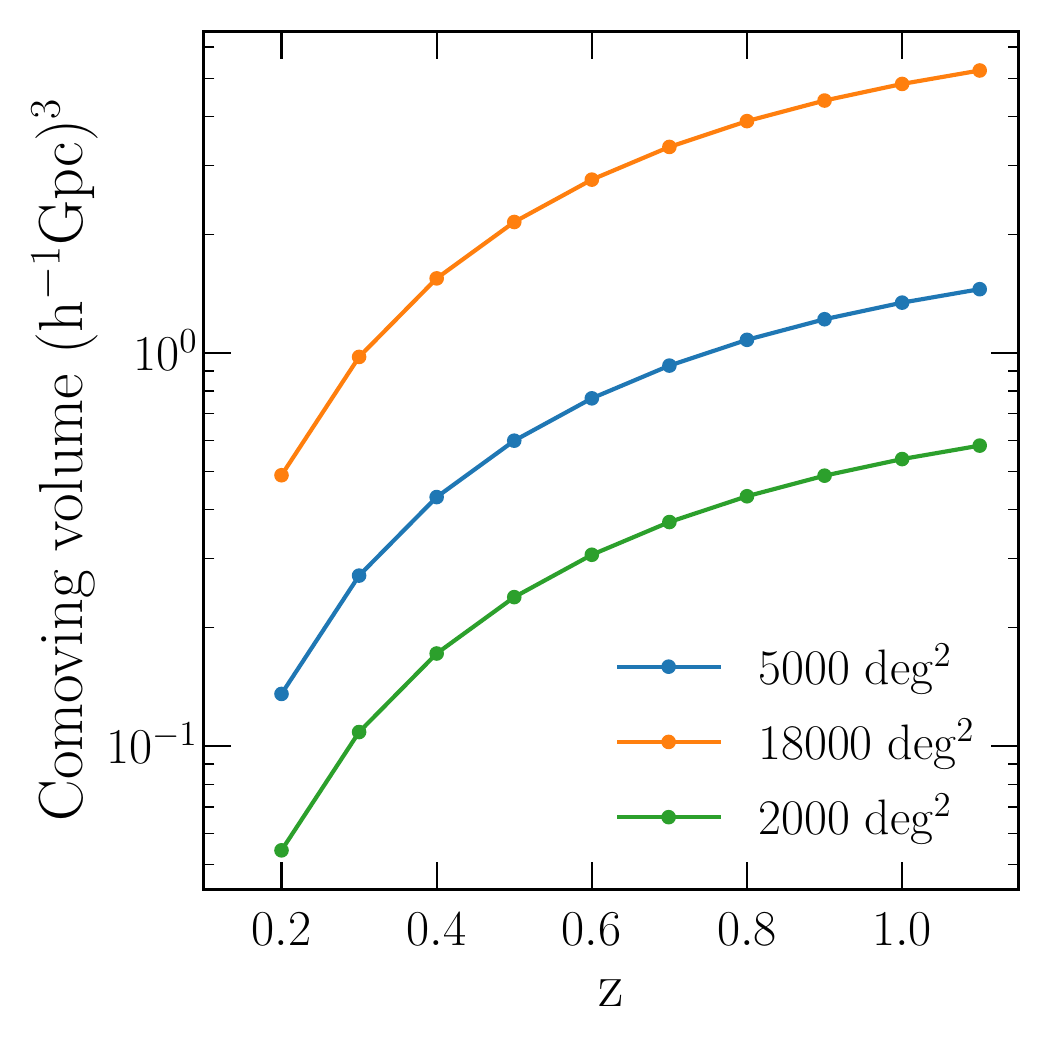}
    \caption{co-moving volume per redshift bin of $\Delta z =0.1$, for various survey areas.}
    \label{fig:survey_volume}
\end{figure}
\begin{figure*}
    \centering
    \includegraphics[width=\textwidth]{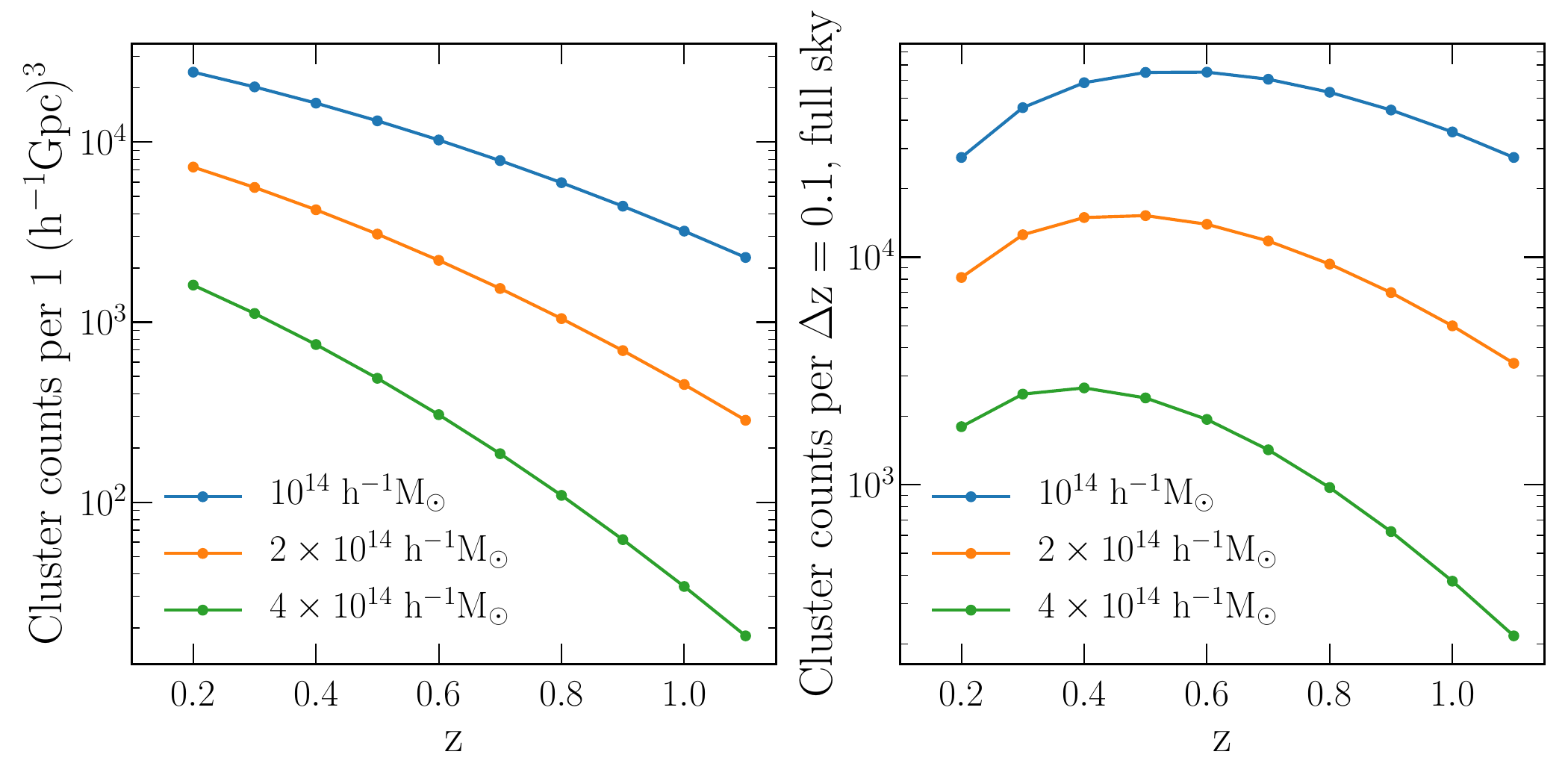}
    \caption{Halo number counts as a function of redshift, for various mass thresholds. 
    {\rm Left}: number counts per 1 $(\hiGpc)^3$
    {\rm Right}: number counts in redshift bins of $\Delta z = 0.1$ in the entire sky.}
    \label{fig:counts}
\end{figure*}

Here we show a few basic numbers for survey volume and cluster counts that are useful for approximate calculations.  Figure~\ref{fig:survey_volume} shows the co-moving volume as a function of redshift, in redshift bins $\Delta z = 0.1$, for three survey areas: 2000, 5000, and 18,000 deg$^2$.  Figure~\ref{fig:counts} shows the halo number counts (zero scatter) above a given mass threshold: the left-hand panel shows the number counts per co-moving volume of 1 $(\hiGpc)^3$, and the right-hand panel shows the number counts per $\Delta z = 0.1$ in the entire sky.

\section{Thresholded vs.\ binned samples}\label{app:bins}

\begin{figure}
    \centering
    \includegraphics[width=0.5\textwidth]{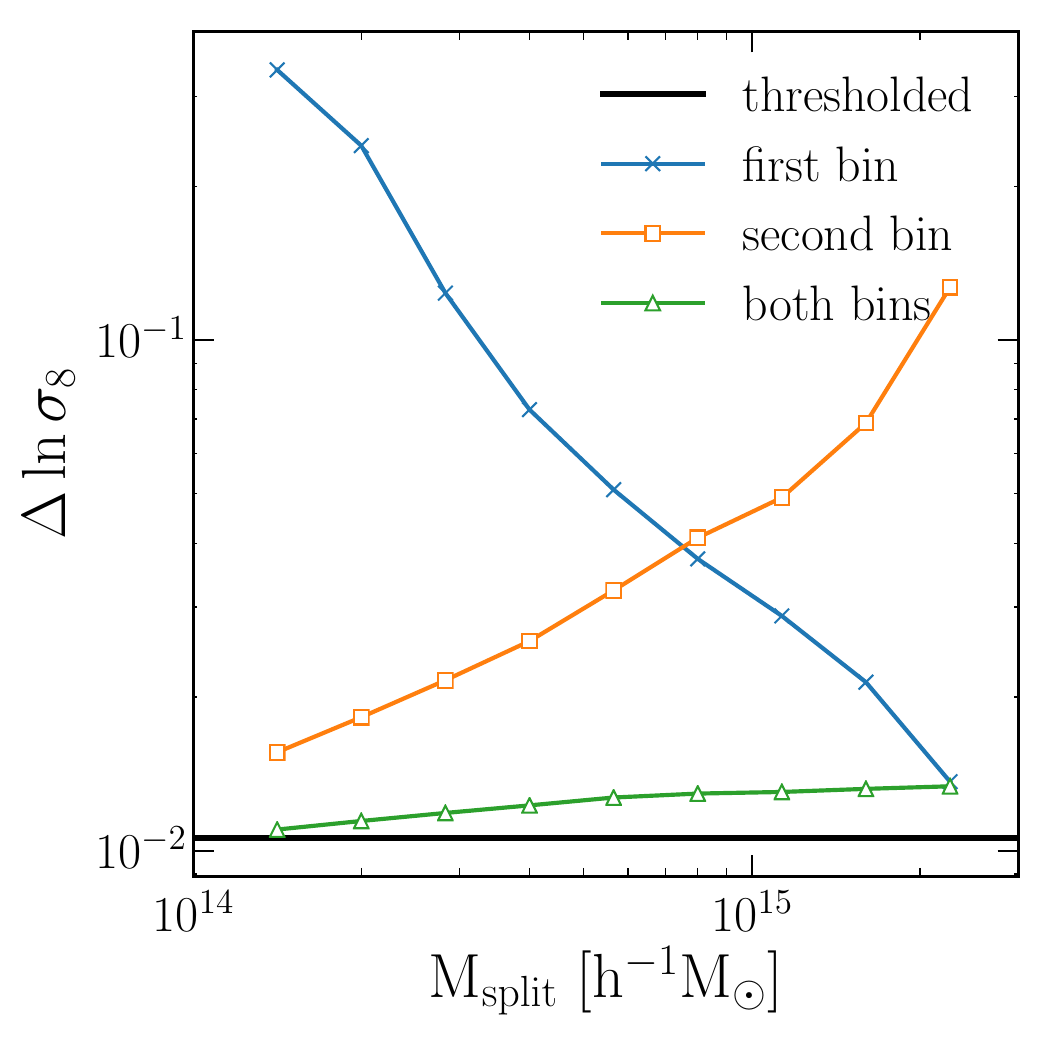}
    \caption{Constraints from thresholded vs.~binned samples.  The black horizontal line corresponds to the baseline assumption $z=0.5$, $\Meqth=10^{14}~\hiMsun$, and $\Delta\sigmalnM=0.1$, with a 1 $(\hiGpc)^3$ survey volume.  We split the sample into two according to the $\Msplit$ (indicated by the $x$-axis).  The blue curve corresponds to the sample below $\Msplit$, and the orange curve corresponds to the sample above $\Msplit$.  The green curve represents the result from combining the two bins.  Splitting the sample into two bins requires extra nuisance parameters and does not present an advantage over using the entire sample.
    }
    \label{fig:constraints_Msplit_2bins}
\end{figure}

In this work we combine all clusters above a threshold of the observable $\Mobsth$.
However, many publications have put clusters into multiple bins of observables \citep[e.g.,][]{DESY1CL}.  In this section, we demonstrate that under the most general assumption for the mass--observable relation (that is, the mass scaling and scatter at the threshold or the bin edges), introducing more bins would lead to more nuisance parameters and present no advantage over a thresholded sample.

We compare the following two cases:
\begin{enumerate}
    \item Case 1: one observable threshold, which corresponds to the (unknown) mass scale $M_1$.  The nuisance parameters are $M_1$ and $\sigma_{\lnM,1}$, with $\sigma_8$ the desired cosmological parameter. 
    \item Case 2: two bins. The first sample corresponds to mass between $M_1$ and $M_2$, and the second sample corresponds to clusters above mass $M_2$.  The nuisance parameters are ($M_1$, $M_2$, $\sigma_{\lnM,1}$, $\sigma_{\lnM,2}$).
\end{enumerate}

For the first case, we only need to constrain the nuisance parameters near the selection threshold, while in the second case, we need to constrain the nuisance parameters at the two selection boundaries.  For the latter, we assume that statistical errors in $\bar{n}$ and $\DS(\rp)$ for the two bins are uncorrelated with each other.  Any correlation between the two samples will further weaken the constraints.

Figure~\ref{fig:constraints_Msplit_2bins} demonstrates the constraints on $\sigma_8$ using one thresholded sample vs.\ two binned samples.
The black horizontal line corresponds to the constraint from the thresholded sample (Case 1).  The blue and orange curves correspond to the $\sigma_8$ constraints from the first and the second bins in Case 2, respectively, and the $x$-axis $\Msplit$ corresponds to where we split the sample into two bins (the changing value of $M_2$).  The first bin (blue curve) has a binning equivalent to $(10^{14}, \ \Msplit)$, and the second bin (orange curves) has a binning equivalent to $(\Msplit, \ \infty)$.

For the largest $\Msplit$, the first bin is equivalent to the thresholded case, and the second bin has little information; for the smallest $\Msplit$, the second bin is equivalent to the thresholded case, and the first bin provides little information.  For the thresholded case, we set the prior on scatter $\Delta\sigmalnM=0.1$; for the two-bin case, we set the prior $\Delta\sigma_{\rm lnM,1} = \Delta\sigma_{\rm lnM,2} = \sqrt{2} \times \Delta\sigmalnM$.  When we combine the two bins, the result is very close to the thresholded case.

When $\Msplit$ is large, $\sigma_{\rm lnM,1}$ in Case 2 plays the same role as $\sigmalnM$ in Case 1, and because we assume $\Delta\sigma_{\rm lnM,1} = \sqrt{2}\Delta\sigmalnM$, Case 2 shows a slightly weaker constraint than a single threshold.  When $\Msplit$ is small, both $\sigma_{\rm lnM,1}$ and $\sigma_{\rm lnM,2}$ in Case 2 describe the scatter near $10^{14}~\hiMsun$, and the combination of the two is equivalent to the prior on $\Delta\sigmalnM$.

Our results show that, if we use the most general assumption for the mass--observable relation, splitting the sample into bins would introduce extra nuisance parameters and does not improve the constraining power.  If one adopts a specific form for the mass--observable relation, such as a power-law mean relation with constant logarithmic scatter, then a binned analysis may appear to yield a stronger cosmological constraint. However, this constraint should be suspect unless the assumed form can be robustly justified.  A useful test for spurious information is to see whether number counts yield a cosmological constraint without calibration of the mean mass--observable relation (e.g., from $\DS$).  Unless the form of $P(\Mobs|M)$ is known almost perfectly, number counts alone have no useful cosmological information, because for any reasonable cosmology, one could reproduce the observed counts by abundance matching.

\section{Analytical estimate of $\sigma_8$ constraints}\label{app:analytic}

\begin{figure*}
    \centering
    \includegraphics[width=\textwidth]{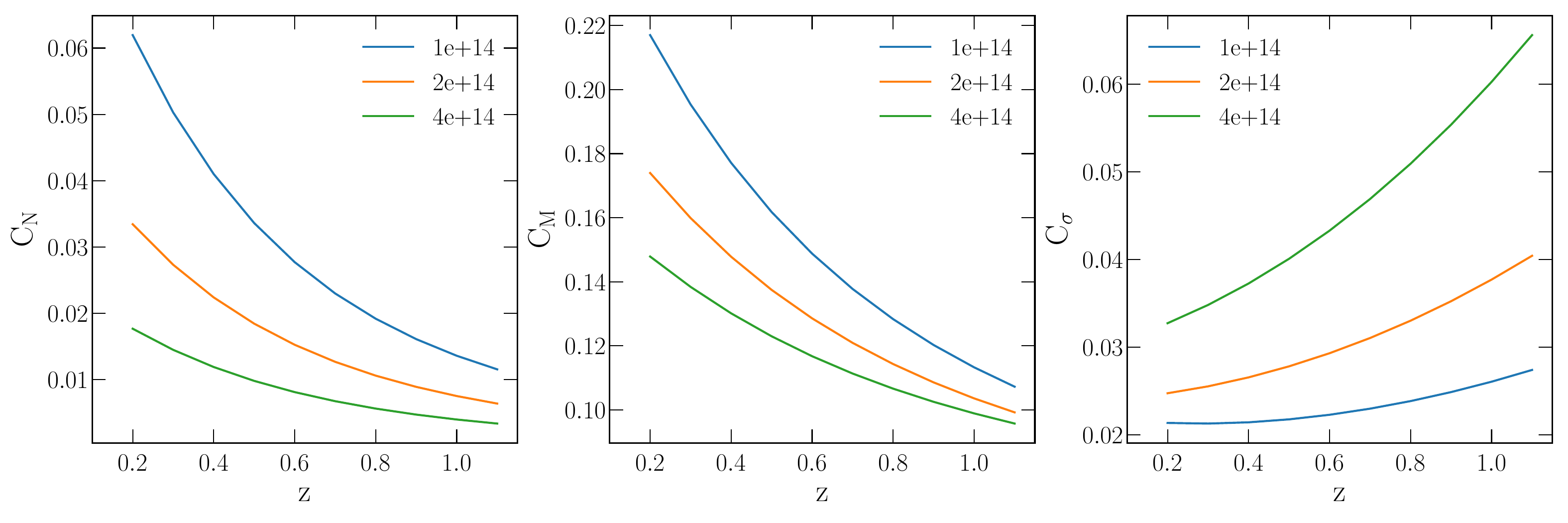}
    \caption{
    Coefficients for analytically calculating the $\sigma_8$ constraints
    (Equation~(\ref{eq:sigma8_analytical})), which can be expressed as the linear combination of the square of the uncertainties of number density, mean mass, and scatter, with the weights given by these coefficients.  The weight on mass uncertainty ($C_M$) is particularly large, indicating the crucial role of mass uncertainties in determining the constraining power of clusters. 
    }
    \label{fig:coefficients}
\end{figure*}
\begin{figure*}
\centering
\includegraphics[width=\textwidth]{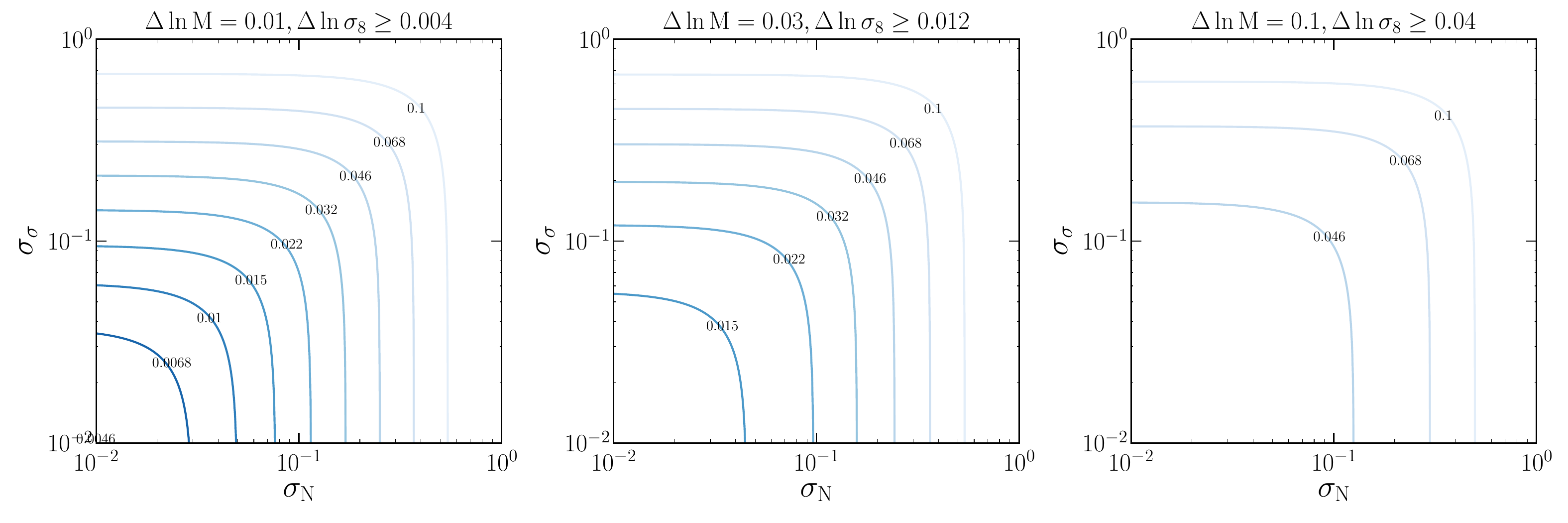}
\caption{Example of trade-off between the error bars on cluster counts, mean mass, and scatter, based on Equation~(\ref{eq:sigma8_analytical}), for $z=0.5$, $10^{14}~\hiMsun$ threshold, and 1 $(\hiGpc)^3$.  Each panel corresponds a fixed error bar in mean mass and a lower limit of $\Delta\lnsigeight$, and different contours correspond to the $\Delta\lnsigeight$ associated with given error bars on number counts ($x$-axis) and scatter ($y$-axis).  For example, if $\Delta\sigmalnM=0.03$ (middle panel), the best possible constraint is $\Delta\lnsigeight=0.012$, but errors of $\sigma_N =0.1$ or $\sigma_\sigma=0.1$ degrade the constraint to $\Delta\lnsigeight=0.022$.
}
\label{fig:constraints_contours}
\end{figure*}
\begin{figure}
    \centering
    \includegraphics[width=0.5\textwidth]{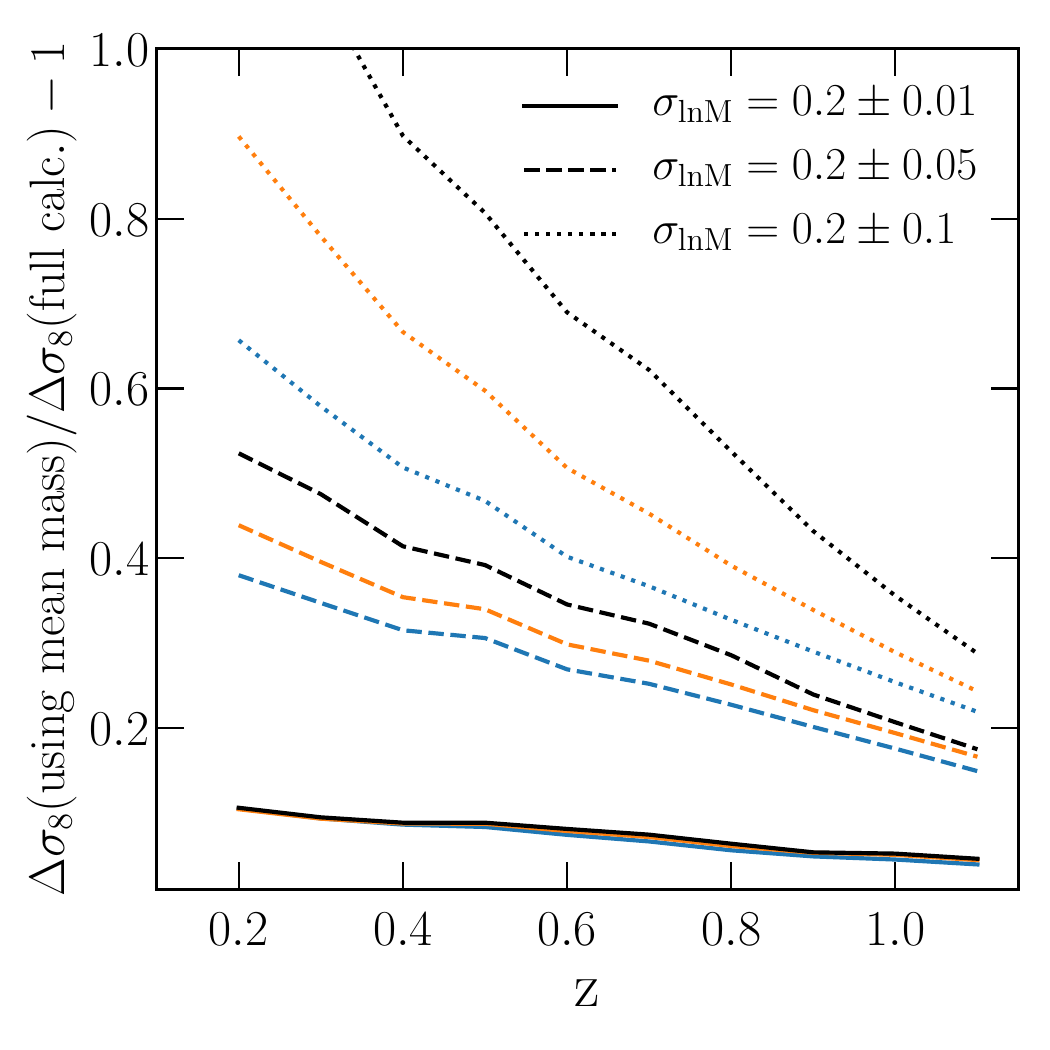}
    \caption{
    Fractional difference between $\sigma_8$ constraints
    calculated using mean mass and using the full lensing profile.  Both cases include the number density information.
    The mean-mass calculation results in slightly weaker constraints than the full lensing calculation, indicating the information lost due to compressing lensing information to mean mass. The difference is larger when the prior on scatter is weaker, but in general the mean-mass calculation is a good approximation and can be easily calculated using the analytical formula (\ref{eq:sigma8_analytical}).
    }
    \label{fig:analytical_accuracy}
\end{figure}

In the main text, we present the constraining power of lensing on the mean mass of the sample.  In this Appendix, we will show that the process of compressing the lensing information from all scales into the constraints on mean mass results in little loss of information for most cases, and one can usually fit the number density and mean mass simultaneously to obtain $\sigma_8$ constraints.  The advantage of the latter approach is that we can estimate the constraints on $\sigma_8$ analytically using the error bars on number density, mean mass, and scatter.  Let us denote
\beqa
N_1 = \frac{d\ln\avgn}{d\lnsigeight} ~,\quad
N_2 = \frac{d\ln\avgn}{d\lnMobsth}  ~,\quad
N_3 = \frac{d\ln\avgn}{d\sigmalnM} ~,\\
M_1 = \frac{d\avg{\lnM}}{d\lnsigeight} ~,\quad
M_2 = \frac{d\avg{\lnM}}{d\lnMobsth} ~,\quad
M_3 = \frac{d\avg{\lnM}}{d\sigmalnM} ~,
\eeqa
and
\beq
\sigma_N = \Delta \ln\avgn   ~,\quad
\sigma_M = \Delta \avg{\lnM}  ~,\quad
\sigma_\sigma = \Delta\sigmalnM   \ .
\eeq
We can calculate the 3$\times$3 Fisher matrix and its inverse analytically.  The constraint on $\lnsigeight$ is given by 
\beqa
(\Delta\lnsigeight)^2 
&= \frac{N_2^2 \sigma_M^2 + M_2^2 \sigma_N^2 + (N_2 M_3 -  M_2 N_3)^2\sigma_\sigma^2}{(N_1 M_2 - M_1 N_2)^2} \\
&= C_N \sigma_N^2 
+ C_M \sigma_M^2 
+ C_\sigma \sigma_\sigma^2    \ .
\label{eq:sigma8_analytical}
\eeqa
That is, $\Delta\lnsigeight$ is the weighted sum of the squares of the uncertainties of number counts, mean mass, and scatter. Figure~\ref{fig:coefficients} shows the coefficients $C_N$, $C_M$, and $C_\sigma$, as a function of redshift ($x$-axis), for various mass thresholds (different lines).   The coefficients $C_M$ are always much larger than $C_N$ and $C_\sigma$, indicating that the uncertainties of mass typically dominate the error budget.  Both $C_N$ and $C_M$ decrease with mass, indicating that high-mass clusters are more sensitive to the reduction of uncertainties.  Nevertheless, since low-mass clusters have much smaller uncertainties, they have the highest constraining power.

Figure~\ref{fig:constraints_contours} shows an example of estimating the constraining power for a survey using this formula.  We use $z=0.5$, $\Meqth=10^{14}~\hiMsun$, and 1 $(\hiGpc)^3$. For each panel, we choose a value of $\sigma_M$, which leads to a lower limit  of $\Delta\lnsigeight$ corresponding to negligible $\sigma_N$ and $\sigma_\sigma$.  In each panel, the contours show $\Delta\lnsigeight$ for given $\sigma_N$ ($x$-axis) and $\sigma_\sigma$ ($y$-axis) values.  Comparing the contours with the same value among these panels shows us the trade-off between these uncertainties.

How well does Equation~(\ref{eq:sigma8_analytical}) work?  Figure~\ref{fig:analytical_accuracy} compares the analytical results (equivalent to compressing all lensing signals into mean-mass constraints) and the full lensing calculation.  The former always gives a slightly larger $\lnsigeight$ error bar, indicating some loss of information due to this compression.  We show the fractional difference between these two approaches, for various levels for shape noise and prior on scatter.  When the scatter is tightly constrained (solid), the difference is less than 10\%.  When the prior on scatter is weak, the analytical formula loses  more constraining power in this data compression.  The derivative of lensing with respect to scatter depends on radius (Figure~\ref{fig:DS}), and when the prior is not well constrained, this scale dependence can provide extra information on scatter compared with using the mean-mass constraints alone.  Figure~\ref{fig:analytical_accuracy} might not look impressive, but it shows that the analytical formula we provide (Equation~(\ref{eq:sigma8_analytical})) can be used for reasonably accurate back-of-the-envelope calculations for estimating the constraining power on $\sigma_8$ of a given survey condition, without computing the full Fisher matrix.  It also allows for comparison of our results to those from other methods of mass calibration, e.g., based on X-ray or SZ observables.  Equation~(\ref{eq:sigma8_analytical}) can also be used to make an approximate estimate when a systematic error on the mean-mass scale, number counts, or mass--observable scatter would influence results at a statistically important level.

\section{Weak correlation between cluster counts and lensing signals}\label{app:abacus_counts_lensing}
\begin{figure}
    \centering
    \includegraphics[width=0.5\textwidth]{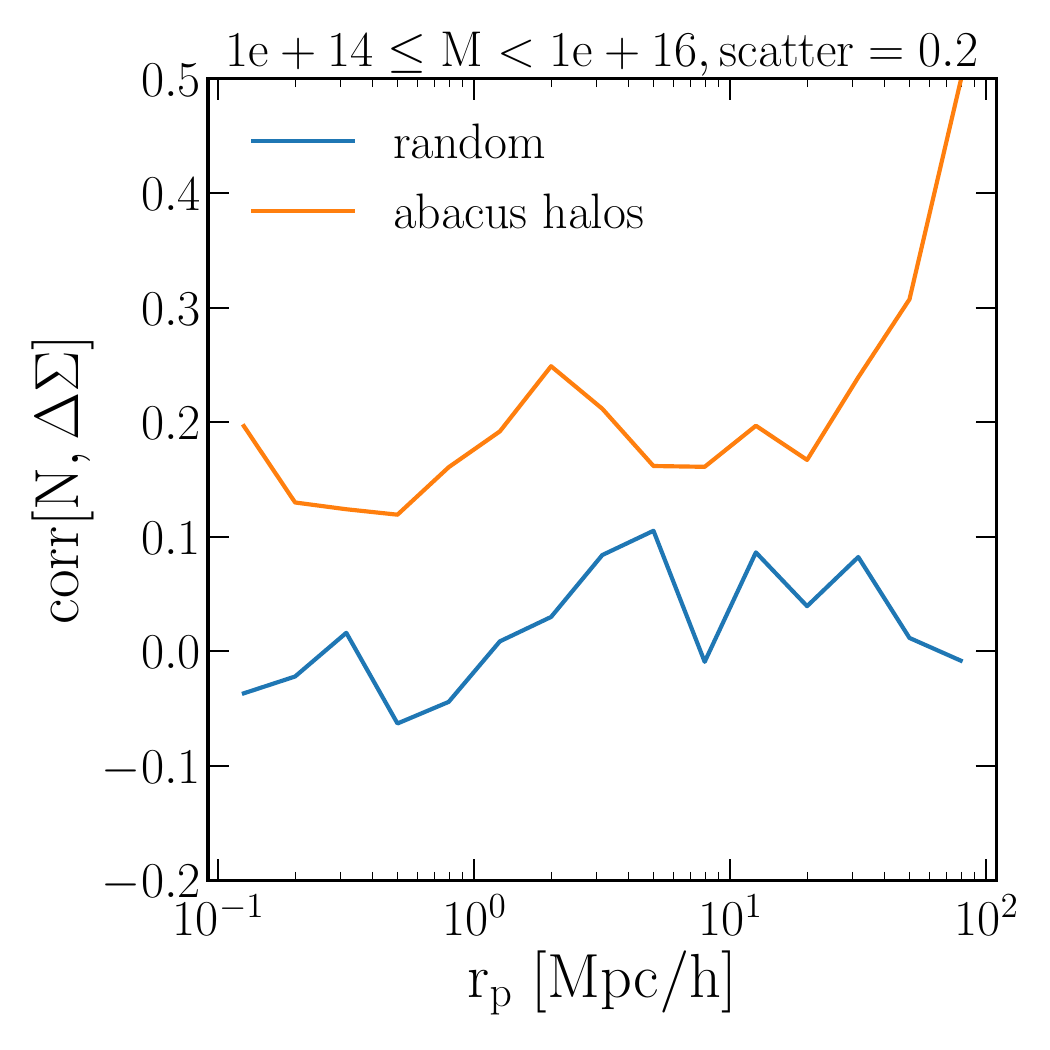}
    \caption{Correlation coefficients between cluster number counts and lensing signal as a function of radius, calculated using the Abacus Cosmos simulations assuming no shape noise.  The orange/blue curves correspond to lensing signals around halos/random points.  The correlation is weak up to 30 $\hiMpc$.}
    \label{fig:abacus_counts_lensing}
\end{figure}
In this work we assume that the covariance between number counts and lensing signals are negligible.  In this Appendix, we show our verification using the Abacus Cosmos simulations \citep{Garrison18}. We measure cluster lensing signals using the 20 boxes of {\tt Abacus\_Cosmos\_720box\_planck} in the same way as in \cite{Wu19}, assuming no shape noise.  We divide each box into nine subvolumes and calculate the number count and lensing signal in each subvolume.  We then measure the correlation coefficients between counts and lensing signals among these 180 realizations.  For comparison, we calculate the lensing signal around random points (20 times the number of clusters).

Figure~\ref{fig:abacus_counts_lensing} shows the correlation coefficient between cluster counts and stacked lensing signals as a function of distance to the cluster center.  The blue curve shows that a 0.1 correlation can arise from the random density fluctuations, and the orange curve shows that the lensing signal around halos has a $\lesssim$ 0.2 correlation with number counts, which is negligible for our purposes.  The correlation increases to approximately 0.5 at $\rp \gtrsim 30\ \hiMpc$, because both halo sample variance and large-scale lensing depend on halo bias (see Equation~\ref{eq:counts_sv}). 
Since $\rp \gtrsim 30\ \hiMpc$ has negligible constraining power (see Figure~\ref{fig:Mmean_rp_split}), ignoring this correlation at large scale does not affect our results.  This figure shows our fiducial mass threshold of $10^{14}\ \hiMsun$; for a higher mass threshold, halo sample variance becomes subdominant, and the correlation will be negligible at all scales.  This calculation also assumes no shape noise for cluster lensing; in real observations, shape noise will further weaken the correlation.

\bibliography{master_refs}{}

\begin{thebibliography}{}
\expandafter\ifx\csname natexlab\endcsname\relax\def\natexlab#1{#1}\fi
\providecommand{\url}[1]{\href{#1}{#1}}
\providecommand{\dodoi}[1]{doi:~\href{http://doi.org/#1}{\nolinkurl{#1}}}
\providecommand{\doeprint}[1]{\href{http://ascl.net/#1}{\nolinkurl{http://ascl.net/#1}}}
\providecommand{\doarXiv}[1]{\href{https://arxiv.org/abs/#1}{\nolinkurl{https://arxiv.org/abs/#1}}}

\bibitem[{{Abbott} {et~al.}(2018){Abbott}, {Abdalla}, {Alarcon}, {Aleksi{\'c}},
  {Allam}, {Allen}, {Amara}, {Annis}, {Asorey}, {Avila}, {Bacon}, {Balbinot},
  {Banerji}, {Banik}, {Barkhouse}, {Baumer}, {Baxter}, {Bechtol}, {Becker},
  {Benoit-L{\'e}vy}, {Benson}, {Bernstein}, {Bertin}, {Blazek}, {Bridle},
  {Brooks}, {Brout}, {Buckley-Geer}, {Burke}, {Busha}, {Campos}, {Capozzi},
  {Carnero Rosell}, {Carrasco Kind}, {Carretero}, {Castander}, {Cawthon},
  {Chang}, {Chen}, {Childress}, {Choi}, {Conselice}, {Crittenden}, {Crocce},
  {Cunha}, {D'Andrea}, {da Costa}, {Das}, {Davis}, {Davis}, {De Vicente},
  {DePoy}, {DeRose}, {Desai}, {Diehl}, {Dietrich}, {Dodelson}, {Doel},
  {Drlica-Wagner}, {Eifler}, {Elliott}, {Elsner}, {Elvin-Poole}, {Estrada},
  {Evrard}, {Fang}, {Fernandez}, {Fert{\'e}}, {Finley}, {Flaugher}, {Fosalba},
  {Friedrich}, {Frieman}, {Garc{\'\i}a-Bellido}, {Garcia-Fernandez}, {Gatti},
  {Gaztanaga}, {Gerdes}, {Giannantonio}, {Gill}, {Glazebrook}, {Goldstein},
  {Gruen}, {Gruendl}, {Gschwend}, {Gutierrez}, {Hamilton}, {Hartley}, {Hinton},
  {Honscheid}, {Hoyle}, {Huterer}, {Jain}, {James}, {Jarvis}, {Jeltema},
  {Johnson}, {Johnson}, {Kacprzak}, {Kent}, {Kim}, {King}, {Kirk}, {Kokron},
  {Kovacs}, {Krause}, {Krawiec}, {Kremin}, {Kuehn}, {Kuhlmann}, {Kuropatkin},
  {Lacasa}, {Lahav}, {Li}, {Liddle}, {Lidman}, {Lima}, {Lin}, {MacCrann},
  {Maia}, {Makler}, {Manera}, {March}, {Marshall}, {Martini}, {McMahon},
  {Melchior}, {Menanteau}, {Miquel}, {Miranda}, {Mudd}, {Muir}, {M{\"o}ller},
  {Neilsen}, {Nichol}, {Nord}, {Nugent}, {Ogando}, {Palmese}, {Peacock},
  {Peiris}, {Peoples}, {Percival}, {Petravick}, {Plazas}, {Porredon}, {Prat},
  {Pujol}, {Rau}, {Refregier}, {Ricker}, {Roe}, {Rollins}, {Romer}, {Roodman},
  {Rosenfeld}, {Ross}, {Rozo}, {Rykoff}, {Sako}, {Salvador}, {Samuroff},
  {S{\'a}nchez}, {Sanchez}, {Santiago}, {Scarpine}, {Schindler}, {Scolnic},
  {Secco}, {Serrano}, {Sevilla-Noarbe}, {Sheldon}, {Smith}, {Smith}, {Smith},
  {Soares-Santos}, {Sobreira}, {Suchyta}, {Tarle}, {Thomas}, {Troxel},
  {Tucker}, {Tucker}, {Uddin}, {Varga}, {Vielzeuf}, {Vikram}, {Vivas},
  {Walker}, {Wang}, {Wechsler}, {Weller}, {Wester}, {Wolf}, {Yanny}, {Yuan},
  {Zenteno}, {Zhang}, {Zhang}, {Zuntz}, \& {Dark Energy Survey
  Collaboration}}]{DESY1KP}
{Abbott}, T.~M.~C., {Abdalla}, F.~B., {Alarcon}, A., {et~al.} 2018, \prd, 98,
  043526, \dodoi{10.1103/PhysRevD.98.043526}

\bibitem[{{Abbott} {et~al.}(2020){Abbott}, {Aguena}, {Alarcon}, {Allam},
  {Allen}, {Annis}, {Avila}, {Bacon}, {Bechtol}, {Bermeo}, {Bernstein},
  {Bertin}, {Bhargava}, {Bocquet}, {Brooks}, {Brout}, {Buckley-Geer}, {Burke},
  {Carnero Rosell}, {Carrasco Kind}, {Carretero}, {Castander}, {Cawthon},
  {Chang}, {Chen}, {Choi}, {Costanzi}, {Crocce}, {da Costa}, {Davis}, {De
  Vicente}, {DeRose}, {Desai}, {Diehl}, {Dietrich}, {Dodelson}, {Doel},
  {Drlica-Wagner}, {Eckert}, {Eifler}, {Elvin-Poole}, {Estrada}, {Everett},
  {Evrard}, {Farahi}, {Ferrero}, {Flaugher}, {Fosalba}, {Frieman},
  {Garc{\'\i}a-Bellido}, {Gatti}, {Gaztanaga}, {Gerdes}, {Giannantonio},
  {Giles}, {Grandis}, {Gruen}, {Gruendl}, {Gschwend}, {Gutierrez}, {Hartley},
  {Hinton}, {Hollowood}, {Honscheid}, {Hoyle}, {Huterer}, {James}, {Jarvis},
  {Jeltema}, {Johnson}, {Johnson}, {Kent}, {Krause}, {Kron}, {Kuehn},
  {Kuropatkin}, {Lahav}, {Li}, {Lidman}, {Lima}, {Lin}, {MacCrann}, {Maia},
  {Mantz}, {Marshall}, {Martini}, {Mayers}, {Melchior}, {Mena-Fern{\'a}ndez},
  {Menanteau}, {Miquel}, {Mohr}, {Nichol}, {Nord}, {Ogand o}, {Palmese},
  {Paz-Chinch{\'o}n}, {Plazas}, {Prat}, {Rau}, {Romer}, {Roodman}, {Rooney},
  {Rozo}, {Rykoff}, {Sako}, {Samuroff}, {S{\'a}nchez}, {Sanchez}, {Saro},
  {Scarpine}, {Schubnell}, {Scolnic}, {Serrano}, {Sevilla-Noarbe}, {Sheldon},
  {Smith}, {Smith}, {Suchyta}, {Swanson}, {Tarle}, {Thomas}, {To}, {Troxel},
  {Tucker}, {Varga}, {von der Linden}, {Walker}, {Wechsler}, {Weller},
  {Wilkinson}, {Wu}, {Yanny}, {Zhang}, {Zhang}, {Zuntz}, \& {DES
  Collaboration}}]{DESY1CL}
{Abbott}, T.~M.~C., {Aguena}, M., {Alarcon}, A., {et~al.} 2020, \prd, 102,
  023509, \dodoi{10.1103/PhysRevD.102.023509}

\bibitem[{{Aguena} \& {Lima}(2018)}]{Aguena18}
{Aguena}, M., \& {Lima}, M. 2018, \prd, 98, 123529,
  \dodoi{10.1103/PhysRevD.98.123529}

\bibitem[{{Akeson} {et~al.}(2019){Akeson}, {Armus}, {Bachelet}, {Bailey},
  {Bartusek}, {Bellini}, {Benford}, {Bennett}, {Bhattacharya}, {Bohlin},
  {Boyer}, {Bozza}, {Bryden}, {Calchi Novati}, {Carpenter}, {Casertano},
  {Choi}, {Content}, {Dayal}, {Dressler}, {Dor{\'e}}, {Fall}, {Fan}, {Fang},
  {Filippenko}, {Finkelstein}, {Foley}, {Furlanetto}, {Kalirai}, {Gaudi},
  {Gilbert}, {Girard}, {Grady}, {Greene}, {Guhathakurta}, {Heinrich},
  {Hemmati}, {Hendel}, {Henderson}, {Henning}, {Hirata}, {Ho}, {Huff},
  {Hutter}, {Jansen}, {Jha}, {Johnson}, {Jones}, {Kasdin}, {Kelly}, {Kirshner},
  {Koekemoer}, {Kruk}, {Lewis}, {Macintosh}, {Madau}, {Malhotra}, {Mand el},
  {Massara}, {Masters}, {McEnery}, {McQuinn}, {Melchior}, {Melton},
  {Mennesson}, {Peeples}, {Penny}, {Perlmutter}, {Pisani}, {Plazas}, {Poleski},
  {Postman}, {Ranc}, {Rauscher}, {Rest}, {Roberge}, {Robertson}, {Rodney},
  {Rhoads}, {Rhodes}, {Ryan}, {Sahu}, {Sand}, {Scolnic}, {Seth}, {Shvartzvald},
  {Siellez}, {Smith}, {Spergel}, {Stassun}, {Street}, {Strolger}, {Szalay},
  {Trauger}, {Troxel}, {Turnbull}, {van der Marel}, {von der Linden}, {Wang},
  {Weinberg}, {Williams}, {Windhorst}, {Wollack}, {Wu}, {Yee}, \&
  {Zimmerman}}]{Akeson19}
{Akeson}, R., {Armus}, L., {Bachelet}, E., {et~al.} 2019, arXiv e-prints,
  arXiv:1902.05569.
\newblock \doarXiv{1902.05569}

\bibitem[{{Allen} {et~al.}(2011){Allen}, {Evrard}, \& {Mantz}}]{Allen11}
{Allen}, S.~W., {Evrard}, A.~E., \& {Mantz}, A.~B. 2011, \araa, 49, 409,
  \dodoi{10.1146/annurev-astro-081710-102514}

\bibitem[{{Anbajagane} {et~al.}(2020){Anbajagane}, {Evrard}, {Farahi},
  {Barnes}, {Dolag}, {McCarthy}, {Nelson}, \& {Pillepich}}]{Anbajagane20}
{Anbajagane}, D., {Evrard}, A.~E., {Farahi}, A., {et~al.} 2020, \mnras, 495,
  686, \dodoi{10.1093/mnras/staa1147}

\bibitem[{{Angulo} {et~al.}(2012){Angulo}, {Springel}, {White}, {Jenkins},
  {Baugh}, \& {Frenk}}]{Angulo12}
{Angulo}, R.~E., {Springel}, V., {White}, S.~D.~M., {et~al.} 2012, \mnras, 426,
  2046, \dodoi{10.1111/j.1365-2966.2012.21830.x}

\bibitem[{{Applegate} {et~al.}(2014){Applegate}, {von der Linden}, {Kelly},
  {Allen}, {Allen}, {Burchat}, {Burke}, {Ebeling}, {Mantz}, \&
  {Morris}}]{Applegate14}
{Applegate}, D.~E., {von der Linden}, A., {Kelly}, P.~L., {et~al.} 2014,
  \mnras, 439, 48, \dodoi{10.1093/mnras/stt2129}

\bibitem[{{Applegate} {et~al.}(2016){Applegate}, {Mantz}, {Allen}, {von der
  Linden}, {Morris}, {Hilbert}, {Kelly}, {Burke}, {Ebeling}, {Rapetti}, \&
  {Schmidt}}]{Applegate16}
{Applegate}, D.~E., {Mantz}, A., {Allen}, S.~W., {et~al.} 2016, \mnras, 457,
  1522, \dodoi{10.1093/mnras/stw005}

\bibitem[{{Baxter} {et~al.}(2016){Baxter}, {Rozo}, {Jain}, {Rykoff}, \&
  {Wechsler}}]{Baxter16}
{Baxter}, E.~J., {Rozo}, E., {Jain}, B., {Rykoff}, E., \& {Wechsler}, R.~H.
  2016, \mnras, 463, 205, \dodoi{10.1093/mnras/stw1939}

\bibitem[{{Benson} {et~al.}(2013){Benson}, {de Haan}, {Dudley}, {Reichardt},
  {Aird}, {Andersson}, {Armstrong}, {Ashby}, {Bautz}, {Bayliss}, {Bazin},
  {Bleem}, {Brodwin}, {Carlstrom}, {Chang}, {Cho}, {Clocchiatti}, {Crawford},
  {Crites}, {Desai}, {Dobbs}, {Foley}, {Forman}, {George}, {Gladders},
  {Gonzalez}, {Halverson}, {Harrington}, {High}, {Holder}, {Holzapfel},
  {Hoover}, {Hrubes}, {Jones}, {Joy}, {Keisler}, {Knox}, {Lee}, {Leitch},
  {Liu}, {Lueker}, {Luong-Van}, {Mantz}, {Marrone}, {McDonald}, {McMahon},
  {Mehl}, {Meyer}, {Mocanu}, {Mohr}, {Montroy}, {Murray}, {Natoli}, {Padin},
  {Plagge}, {Pryke}, {Rest}, {Ruel}, {Ruhl}, {Saliwanchik}, {Saro}, {Sayre},
  {Schaffer}, {Shaw}, {Shirokoff}, {Song}, {Spieler}, {Stalder},
  {Staniszewski}, {Stark}, {Story}, {Stubbs}, {Suhada}, {van Engelen},
  {Vanderlinde}, {Vieira}, {Vikhlinin}, {Williamson}, {Zahn}, \&
  {Zenteno}}]{Benson13}
{Benson}, B.~A., {de Haan}, T., {Dudley}, J.~P., {et~al.} 2013, \apj, 763, 147,
  \dodoi{10.1088/0004-637X/763/2/147}

\bibitem[{{Bleem} {et~al.}(2020){Bleem}, {Bocquet}, {Stalder}, {Gladders},
  {Ade}, {Allen}, {Anderson}, {Annis}, {Ashby}, {Austermann}, {Avila}, {Avva},
  {Bayliss}, {Beall}, {Bechtol}, {Bender}, {Benson}, {Bertin}, {Bianchini},
  {Blake}, {Brodwin}, {Brooks}, {Buckley-Geer}, {Burke}, {Carlstrom}, {Rosell},
  {Carrasco Kind}, {Carretero}, {Chang}, {Chiang}, {Citron}, {Moran},
  {Costanzi}, {Crawford}, {Crites}, {da Costa}, {de Haan}, {De Vicente},
  {Desai}, {Diehl}, {Dietrich}, {Dobbs}, {Eifler}, {Everett}, {Flaugher},
  {Floyd}, {Frieman}, {Gallicchio}, {Garc{\'\i}a-Bellido}, {George}, {Gerdes},
  {Gilbert}, {Gruen}, {Gruendl}, {Gschwend}, {Gupta}, {Gutierrez}, {Halverson},
  {Harrington}, {Henning}, {Heymans}, {Holder}, {Hollowood}, {Holzapfel},
  {Honscheid}, {Hrubes}, {Huang}, {Hubmayr}, {Irwin}, {James}, {Jeltema},
  {Joudaki}, {Khullar}, {Klein}, {Knox}, {Kuropatkin}, {Lee}, {Li}, {Lidman},
  {Lowitz}, {MacCrann}, {Mahler}, {Maia}, {Marshall}, {McDonald}, {McMahon},
  {Melchior}, {Menanteau}, {Meyer}, {Miquel}, {Mocanu}, {Mohr}, {Montgomery},
  {Nadolski}, {Natoli}, {Nibarger}, {Noble}, {Novosad}, {Padin}, {Palmese},
  {Parkinson}, {Patil}, {Paz-Chinch{\'o}n}, {Plazas}, {Pryke}, {Ramachandra},
  {Reichardt}, {Remolina Gonz{\'a}lez}, {Romer}, {Roodman}, {Ruhl}, {Rykoff},
  {Saliwanchik}, {Sanchez}, {Saro}, {Sayre}, {Schaffer}, {Schrabback},
  {Serrano}, {Sharon}, {Sievers}, {Smecher}, {Smith}, {Soares-Santos}, {Stark},
  {Story}, {Suchyta}, {Tarle}, {Tucker}, {Vanderlinde}, {Veach}, {Vieira},
  {Wang}, {Weller}, {Whitehorn}, {Wu}, {Yefremenko}, \& {Zhang}}]{Bleem20}
{Bleem}, L.~E., {Bocquet}, S., {Stalder}, B., {et~al.} 2020, \apjs, 247, 25,
  \dodoi{10.3847/1538-4365/ab6993}

\bibitem[{{Bocquet} {et~al.}(2019){Bocquet}, {Dietrich}, {Schrabback}, {Bleem},
  {Klein}, {Allen}, {Applegate}, {Ashby}, {Bautz}, {Bayliss}, {Benson},
  {Brodwin}, {Bulbul}, {Canning}, {Capasso}, {Carlstrom}, {Chang}, {Chiu},
  {Cho}, {Clocchiatti}, {Crawford}, {Crites}, {de Haan}, {Desai}, {Dobbs},
  {Foley}, {Forman}, {Garmire}, {George}, {Gladders}, {Gonzalez}, {Grandis},
  {Gupta}, {Halverson}, {Hlavacek-Larrondo}, {Hoekstra}, {Holder}, {Holzapfel},
  {Hou}, {Hrubes}, {Huang}, {Jones}, {Khullar}, {Knox}, {Kraft}, {Lee}, {von
  der Linden}, {Luong-Van}, {Mantz}, {Marrone}, {McDonald}, {McMahon}, {Meyer},
  {Mocanu}, {Mohr}, {Morris}, {Padin}, {Patil}, {Pryke}, {Rapetti},
  {Reichardt}, {Rest}, {Ruhl}, {Saliwanchik}, {Saro}, {Sayre}, {Schaffer},
  {Shirokoff}, {Stalder}, {Stanford}, {Staniszewski}, {Stark}, {Story},
  {Strazzullo}, {Stubbs}, {Vanderlinde}, {Vieira}, {Vikhlinin}, {Williamson},
  \& {Zenteno}}]{Bocquet19}
{Bocquet}, S., {Dietrich}, J.~P., {Schrabback}, T., {et~al.} 2019, \apj, 878,
  55, \dodoi{10.3847/1538-4357/ab1f10}

\bibitem[{{Bradshaw} {et~al.}(2020){Bradshaw}, {Leauthaud}, {Hearin}, {Huang},
  \& {Behroozi}}]{Bradshaw20}
{Bradshaw}, C., {Leauthaud}, A., {Hearin}, A., {Huang}, S., \& {Behroozi}, P.
  2020, \mnras, 493, 337, \dodoi{10.1093/mnras/staa081}

\bibitem[{{Chiu} {et~al.}(2020){Chiu}, {Okumura}, {Oguri}, {Agrawal}, {Umetsu},
  \& {Lin}}]{Chiu20}
{Chiu}, I.~N., {Okumura}, T., {Oguri}, M., {et~al.} 2020, \mnras, 498, 2030,
  \dodoi{10.1093/mnras/staa2440}

\bibitem[{{Correa} {et~al.}(2015){Correa}, {Wyithe}, {Schaye}, \&
  {Duffy}}]{Correa15}
{Correa}, C.~A., {Wyithe}, J. S.~B., {Schaye}, J., \& {Duffy}, A.~R. 2015,
  \mnras, 452, 1217, \dodoi{10.1093/mnras/stv1363}

\bibitem[{{Costanzi} {et~al.}(2019){Costanzi}, {Rozo}, {Simet}, {Zhang},
  {Evrard}, {Mantz}, {Rykoff}, {Jeltema}, {Gruen}, {Allen}, {McClintock},
  {Romer}, {von der Linden}, {Farahi}, {DeRose}, {Varga}, {Weller}, {Giles},
  {Hollowood}, {Bhargava}, {Bermeo-Hernandez}, {Chen}, {Abbott}, {Abdalla},
  {Avila}, {Bechtol}, {Brooks}, {Buckley-Geer}, {Burke}, {Rosell}, {Kind},
  {Carretero}, {Crocce}, {Cunha}, {da Costa}, {Davis}, {De Vicente}, {Diehl},
  {Dietrich}, {Doel}, {Eifler}, {Estrada}, {Flaugher}, {Fosalba}, {Frieman},
  {Garc{\'\i}a-Bellido}, {Gaztanaga}, {Gerdes}, {Giannantonio}, {Gruendl},
  {Gschwend}, {Gutierrez}, {Hartley}, {Honscheid}, {Hoyle}, {James}, {Krause},
  {Kuehn}, {Kuropatkin}, {Lima}, {Lin}, {Maia}, {March}, {Marshall}, {Martini},
  {Menanteau}, {Miller}, {Miquel}, {Mohr}, {Ogando}, {Plazas}, {Roodman},
  {Sanchez}, {Scarpine}, {Schindler}, {Schubnell}, {Serrano}, {Sevilla-Noarbe},
  {Sheldon}, {Smith}, {Soares-Santos}, {Sobreira}, {Suchyta}, {Swanson},
  {Tarle}, {Thomas}, \& {Wechsler}}]{Costanzi19SDSS}
{Costanzi}, M., {Rozo}, E., {Simet}, M., {et~al.} 2019, \mnras, 488, 4779,
  \dodoi{10.1093/mnras/stz1949}

\bibitem[{{Dietrich} {et~al.}(2014){Dietrich}, {Zhang}, {Song}, {Davis},
  {McKay}, {Baruah}, {Becker}, {Benoist}, {Busha}, {da Costa}, {Hao}, {Maia},
  {Miller}, {Ogando}, {Romer}, {Rozo}, {Rykoff}, \& {Wechsler}}]{Dietrich14}
{Dietrich}, J.~P., {Zhang}, Y., {Song}, J., {et~al.} 2014, Monthly Notices of
  the Royal Astronomical Society, 443, 1713, \dodoi{10.1093/mnras/stu1282}

\bibitem[{{Eifler} {et~al.}(2020{\natexlab{a}}){Eifler}, {Miyatake}, {Krause},
  {Heinrich}, {Miranda}, {Hirata}, {Xu}, {Hemmati}, {Simet}, {Capak}, {Choi},
  {Dore}, {Doux}, {Fang}, {Hounsell}, {Huff}, {Huang}, {Jarvis}, {Masters},
  {Rozo}, {Scolnic}, {Spergel}, {Troxel}, {von der Linden}, {Wang}, {Weinberg},
  {Wenzl}, \& {Wu}}]{Eifler20HLS}
{Eifler}, T., {Miyatake}, H., {Krause}, E., {et~al.} 2020{\natexlab{a}}, arXiv
  e-prints, arXiv:2004.05271.
\newblock \doarXiv{2004.05271}

\bibitem[{{Eifler} {et~al.}(2020{\natexlab{b}}){Eifler}, {Simet}, {Krause},
  {Hirata}, {Huang}, {Fang}, {Mirand a}, {Mandelbaum}, {Doux}, {Heinrich},
  {Huff}, {Miyatake}, {Hemmati}, {Xu}, {Rogozenski}, {Capak}, {Choi}, {Dore},
  {Jain}, {Jarvis}, {MacCrann}, {Masters}, {Rozo}, {Spergel}, {Troxel}, {von
  der Linden}, {Wang}, {Weinberg}, {Wenzl}, \& {Wu}}]{Eifler20WFIRSTLSST}
{Eifler}, T., {Simet}, M., {Krause}, E., {et~al.} 2020{\natexlab{b}}, arXiv
  e-prints, arXiv:2004.04702.
\newblock \doarXiv{2004.04702}

\bibitem[{{Evrard}(1989)}]{Evrard89}
{Evrard}, A.~E. 1989, \apjl, 341, L71, \dodoi{10.1086/185460}

\bibitem[{{Farahi} {et~al.}(2019){Farahi}, {Chen}, {Evrard}, {Hollowood},
  {Wilkinson}, {Bhargava}, {Giles}, {Romer}, {Jeltema}, {Hilton}, {Bermeo},
  {Mayers}, {Vergara Cervantes}, {Rozo}, {Rykoff}, {Collins}, {Costanzi},
  {Everett}, {Liddle}, {Mann}, {Mantz}, {Rooney}, {Sahlen}, {Stott}, {Viana},
  {Zhang}, {Annis}, {Avila}, {Brooks}, {Buckley-Geer}, {Burke}, {Carnero
  Rosell}, {Carrasco Kind}, {Carretero}, {Castander}, {da Costa}, {De Vicente},
  {Desai}, {Diehl}, {Dietrich}, {Doel}, {Flaugher}, {Fosalba}, {Frieman},
  {Garc{\'\i}a-Bellido}, {Gaztanaga}, {Gerdes}, {Gruen}, {Gruendl}, {Gschwend},
  {Gutierrez}, {Honscheid}, {James}, {Krause}, {Kuehn}, {Kuropatkin}, {Lima},
  {Maia}, {Marshall}, {Melchior}, {Menanteau}, {Miquel}, {Ogando}, {Plazas},
  {Sanchez}, {Scarpine}, {Schubnell}, {Serrano}, {Sevilla-Noarbe}, {Smith},
  {Sobreira}, {Suchyta}, {Swanson}, {Tarle}, {Thomas}, {Tucker}, {Vikram},
  {Walker}, {Weller}, \& {DES Collaboration}}]{Farahi19}
{Farahi}, A., {Chen}, X., {Evrard}, A.~E., {et~al.} 2019, \mnras, 490, 3341,
  \dodoi{10.1093/mnras/stz2689}

\bibitem[{{Garrison} {et~al.}(2018){Garrison}, {Eisenstein}, {Ferrer},
  {Tinker}, {Pinto}, \& {Weinberg}}]{Garrison18}
{Garrison}, L.~H., {Eisenstein}, D.~J., {Ferrer}, D., {et~al.} 2018, \apjs,
  236, 43, \dodoi{10.3847/1538-4365/aabfd3}

\bibitem[{{Hahn} {et~al.}(2017){Hahn}, {Martizzi}, {Wu}, {Evrard}, {Teyssier},
  \& {Wechsler}}]{Hahn17}
{Hahn}, O., {Martizzi}, D., {Wu}, H.-Y., {et~al.} 2017, \mnras, 470, 166,
  \dodoi{10.1093/mnras/stx001}

\bibitem[{{Haiman} {et~al.}(2001){Haiman}, {Mohr}, \& {Holder}}]{Haiman01}
{Haiman}, Z., {Mohr}, J.~J., \& {Holder}, G.~P. 2001, \apj, 553, 545,
  \dodoi{10.1086/320939}

\bibitem[{{Hayashi} \& {White}(2008)}]{HayashiWhite08}
{Hayashi}, E., \& {White}, S.~D.~M. 2008, \mnras, 388, 2,
  \dodoi{10.1111/j.1365-2966.2008.13371.x}

\bibitem[{{Hemmati} {et~al.}(2019){Hemmati}, {Capak}, {Masters}, {Davidzon},
  {Dor{\`e}}, {Kruk}, {Mobasher}, {Rhodes}, {Scolnic}, \& {Stern}}]{Hemmati19}
{Hemmati}, S., {Capak}, P., {Masters}, D., {et~al.} 2019, \apj, 877, 117,
  \dodoi{10.3847/1538-4357/ab1be5}

\bibitem[{{Henson} {et~al.}(2017){Henson}, {Barnes}, {Kay}, {McCarthy}, \&
  {Schaye}}]{Henson17}
{Henson}, M.~A., {Barnes}, D.~J., {Kay}, S.~T., {McCarthy}, I.~G., \& {Schaye},
  J. 2017, \mnras, 465, 3361, \dodoi{10.1093/mnras/stw2899}

\bibitem[{{Herbonnet} {et~al.}(2019){Herbonnet}, {von der Linden}, {Allen},
  {Mantz}, {Modumudi}, {Morris}, \& {Kelly}}]{Herbonnet19}
{Herbonnet}, R., {von der Linden}, A., {Allen}, S.~W., {et~al.} 2019, \mnras,
  490, 4889, \dodoi{10.1093/mnras/stz2913}

\bibitem[{{Holder} {et~al.}(2001){Holder}, {Haiman}, \& {Mohr}}]{Holder01}
{Holder}, G.~P., {Haiman}, Z., \& {Mohr}, J.~J. 2001, \apjl, 560, L111,
  \dodoi{10.1086/324309}

\bibitem[{{Hu} \& {Kravtsov}(2003)}]{HuKravtsov03}
{Hu}, W., \& {Kravtsov}, A.~V. 2003, \apj, 584, 702, \dodoi{10.1086/345846}

\bibitem[{{Kilbinger}(2015)}]{Kilbinger15}
{Kilbinger}, M. 2015, Reports on Progress in Physics, 78, 086901,
  \dodoi{10.1088/0034-4885/78/8/086901}

\bibitem[{{Kravtsov} \& {Borgani}(2012)}]{KravtsovBorgani12}
{Kravtsov}, A.~V., \& {Borgani}, S. 2012, \araa, 50, 353,
  \dodoi{10.1146/annurev-astro-081811-125502}

\bibitem[{{Kravtsov} {et~al.}(2006){Kravtsov}, {Vikhlinin}, \&
  {Nagai}}]{Kravtsov06}
{Kravtsov}, A.~V., {Vikhlinin}, A., \& {Nagai}, D. 2006, \apj, 650, 128,
  \dodoi{10.1086/506319}

\bibitem[{{Le Brun} {et~al.}(2017){Le Brun}, {McCarthy}, {Schaye}, \&
  {Ponman}}]{LeBrun17}
{Le Brun}, A. M.~C., {McCarthy}, I.~G., {Schaye}, J., \& {Ponman}, T.~J. 2017,
  \mnras, 466, 4442, \dodoi{10.1093/mnras/stw3361}

\bibitem[{{Mandelbaum} {et~al.}(2006){Mandelbaum}, {Seljak}, {Cool}, {Blanton},
  {Hirata}, \& {Brinkmann}}]{Mandelbaum06}
{Mandelbaum}, R., {Seljak}, U., {Cool}, R.~J., {et~al.} 2006, \mnras, 372, 758,
  \dodoi{10.1111/j.1365-2966.2006.10906.x}

\bibitem[{{Mandelbaum} {et~al.}(2013){Mandelbaum}, {Slosar}, {Baldauf},
  {Seljak}, {Hirata}, {Nakajima}, {Reyes}, \& {Smith}}]{Mandelbaum13}
{Mandelbaum}, R., {Slosar}, A., {Baldauf}, T., {et~al.} 2013, \mnras, 432,
  1544, \dodoi{10.1093/mnras/stt572}

\bibitem[{{Mandelbaum} {et~al.}(2005){Mandelbaum}, {Hirata}, {Seljak}, {Guzik},
  {Padmanabhan}, {Blake}, {Blanton}, {Lupton}, \& {Brinkmann}}]{Mandelbaum05}
{Mandelbaum}, R., {Hirata}, C.~M., {Seljak}, U., {et~al.} 2005, \mnras, 361,
  1287, \dodoi{10.1111/j.1365-2966.2005.09282.x}

\bibitem[{{Mantz} {et~al.}(2010){Mantz}, {Allen}, {Rapetti}, \&
  {Ebeling}}]{Mantz10}
{Mantz}, A., {Allen}, S.~W., {Rapetti}, D., \& {Ebeling}, H. 2010, \mnras, 406,
  1759, \dodoi{10.1111/j.1365-2966.2010.16992.x}

\bibitem[{{Mantz} {et~al.}(2014){Mantz}, {Allen}, {Morris}, {Rapetti},
  {Applegate}, {Kelly}, {von der Linden}, \& {Schmidt}}]{Mantz14}
{Mantz}, A.~B., {Allen}, S.~W., {Morris}, R.~G., {et~al.} 2014, \mnras, 440,
  2077, \dodoi{10.1093/mnras/stu368}

\bibitem[{{McClintock} {et~al.}(2019){McClintock}, {Varga}, {Gruen}, {Rozo},
  {Rykoff}, {Shin}, {Melchior}, {DeRose}, {Seitz}, {Dietrich}, {Sheldon},
  {Zhang}, {von der Linden}, {Jeltema}, {Mantz}, {Romer}, {Allen}, {Becker},
  {Bermeo}, {Bhargava}, {Costanzi}, {Everett}, {Farahi}, {Hamaus}, {Hartley},
  {Hollowood}, {Hoyle}, {Israel}, {Li}, {MacCrann}, {Morris}, {Palmese},
  {Plazas}, {Pollina}, {Rau}, {Simet}, {Soares-Santos}, {Troxel}, {Vergara
  Cervantes}, {Wechsler}, {Zuntz}, {Abbott}, {Abdalla}, {Allam}, {Annis},
  {Avila}, {Bridle}, {Brooks}, {Burke}, {Carnero Rosell}, {Carrasco Kind},
  {Carretero}, {Castander}, {Crocce}, {Cunha}, {D'Andrea}, {da Costa}, {Davis},
  {De Vicente}, {Diehl}, {Doel}, {Drlica-Wagner}, {Evrard}, {Flaugher},
  {Fosalba}, {Frieman}, {Garc{\'{\i}}a-Bellido}, {Gaztanaga}, {Gerdes},
  {Giannantonio}, {Gruendl}, {Gutierrez}, {Honscheid}, {James}, {Kirk},
  {Krause}, {Kuehn}, {Lahav}, {Li}, {Lima}, {March}, {Marshall}, {Menanteau},
  {Miquel}, {Mohr}, {Nord}, {Ogando}, {Roodman}, {Sanchez}, {Scarpine},
  {Schindler}, {Sevilla-Noarbe}, {Smith}, {Smith}, {Sobreira}, {Suchyta},
  {Swanson}, {Tarle}, {Tucker}, {Vikram}, {Walker}, \& {Weller}}]{McClintock19}
{McClintock}, T., {Varga}, T.~N., {Gruen}, D., {et~al.} 2019, \mnras, 482,
  1352, \dodoi{10.1093/mnras/sty2711}

\bibitem[{{McKay} {et~al.}(2001){McKay}, {Sheldon}, {Racusin}, {Fischer},
  {Seljak}, {Stebbins}, {Johnston}, {Frieman}, {Bahcall}, {Brinkmann},
  {Csabai}, {Fukugita}, {Hennessy}, {Ivezic}, {Lamb}, {Loveday}, {Lupton},
  {Munn}, {Nichol}, {Pier}, \& {York}}]{McKay01}
{McKay}, T.~A., {Sheldon}, E.~S., {Racusin}, J., {et~al.} 2001, arXiv e-prints,
  astro-ph/0108013.
\newblock \doarXiv{astro-ph/0108013}

\bibitem[{{Mead} {et~al.}(2020){Mead}, {Tr{\"o}ster}, {Heymans}, {Van
  Waerbeke}, \& {McCarthy}}]{Mead20}
{Mead}, A.~J., {Tr{\"o}ster}, T., {Heymans}, C., {Van Waerbeke}, L., \&
  {McCarthy}, I.~G. 2020, \aap, 641, A130, \dodoi{10.1051/0004-6361/202038308}

\bibitem[{{Melchior} {et~al.}(2017){Melchior}, {Gruen}, {McClintock}, {Varga},
  {Sheldon}, {Rozo}, {Amara}, {Becker}, {Benson}, {Bermeo}, {Bridle},
  {Clampitt}, {Dietrich}, {Hartley}, {Hollowood}, {Jain}, {Jarvis}, {Jeltema},
  {Kacprzak}, {MacCrann}, {Rykoff}, {Saro}, {Suchyta}, {Troxel}, {Zuntz},
  {Bonnett}, {Plazas}, {Abbott}, {Abdalla}, {Annis}, {Benoit-L{\'e}vy},
  {Bernstein}, {Bertin}, {Brooks}, {Buckley-Geer}, {Carnero Rosell}, {Carrasco
  Kind}, {Carretero}, {Cunha}, {D'Andrea}, {da Costa}, {Desai}, {Eifler},
  {Flaugher}, {Fosalba}, {Garc{\'{\i}}a-Bellido}, {Gaztanaga}, {Gerdes},
  {Gruendl}, {Gschwend}, {Gutierrez}, {Honscheid}, {James}, {Kirk}, {Krause},
  {Kuehn}, {Kuropatkin}, {Lahav}, {Lima}, {Maia}, {March}, {Martini},
  {Menanteau}, {Miller}, {Miquel}, {Mohr}, {Nichol}, {Ogando}, {Romer},
  {Sanchez}, {Scarpine}, {Sevilla-Noarbe}, {Smith}, {Soares-Santos},
  {Sobreira}, {Swanson}, {Tarle}, {Thomas}, {Walker}, {Weller}, \&
  {Zhang}}]{Melchior17}
{Melchior}, P., {Gruen}, D., {McClintock}, T., {et~al.} 2017, \mnras, 469,
  4899, \dodoi{10.1093/mnras/stx1053}

\bibitem[{{Miller} {et~al.}(2001){Miller}, {Nichol}, \& {Batuski}}]{Miller01}
{Miller}, C.~J., {Nichol}, R.~C., \& {Batuski}, D.~J. 2001, \apj, 555, 68,
  \dodoi{10.1086/321468}

\bibitem[{{Murata} {et~al.}(2019){Murata}, {Oguri}, {Nishimichi}, {Takada},
  {Mandelbaum}, {More}, {Shirasaki}, {Nishizawa}, \& {Osato}}]{Murata19}
{Murata}, R., {Oguri}, M., {Nishimichi}, T., {et~al.} 2019, \pasj, 71, 107,
  \dodoi{10.1093/pasj/psz092}

\bibitem[{{Nagai}(2006)}]{Nagai06}
{Nagai}, D. 2006, \apj, 650, 538, \dodoi{10.1086/506467}

\bibitem[{{Navarro} {et~al.}(1997){Navarro}, {Frenk}, \& {White}}]{NFW97}
{Navarro}, J.~F., {Frenk}, C.~S., \& {White}, S.~D.~M. 1997, \apj, 490, 493,
  \dodoi{10.1086/304888}

\bibitem[{{Oguri} \& {Takada}(2011)}]{OguriTakada11}
{Oguri}, M., \& {Takada}, M. 2011, \prd, 83, 023008,
  \dodoi{10.1103/PhysRevD.83.023008}

\bibitem[{{Ohio Supercomputer Center}(1987)}]{OSC}
{Ohio Supercomputer Center}. 1987, Ohio Supercomputer Center.
\newblock \url{http://osc.edu/ark:/19495/f5s1ph73}

\bibitem[{{Osato} {et~al.}(2018){Osato}, {Nishimichi}, {Oguri}, {Takada}, \&
  {Okumura}}]{Osato18}
{Osato}, K., {Nishimichi}, T., {Oguri}, M., {Takada}, M., \& {Okumura}, T.
  2018, Monthly Notices of the Royal Astronomical Society, 477, 2141,
  \dodoi{10.1093/mnras/sty762}

\bibitem[{{Peebles} {et~al.}(1989){Peebles}, {Daly}, \&
  {Juszkiewicz}}]{Peebles89}
{Peebles}, P.~J.~E., {Daly}, R.~A., \& {Juszkiewicz}, R. 1989, \apj, 347, 563,
  \dodoi{10.1086/168149}

\bibitem[{{Pillepich} {et~al.}(2018){Pillepich}, {Nelson}, {Hernquist},
  {Springel}, {Pakmor}, {Torrey}, {Weinberger}, {Genel}, {Naiman}, {Marinacci},
  \& {Vogelsberger}}]{Pillepich18sm}
{Pillepich}, A., {Nelson}, D., {Hernquist}, L., {et~al.} 2018, \mnras, 475,
  648, \dodoi{10.1093/mnras/stx3112}

\bibitem[{{Planck Collaboration} {et~al.}(2014){Planck Collaboration}, {Ade},
  {Aghanim}, {Armitage-Caplan}, {Arnaud}, {Ashdown}, {Atrio-Barandela},
  {Aumont}, {Baccigalupi}, {Banday}, {Barreiro}, {Barrena}, {Bartlett},
  {Battaner}, {Battye}, {Benabed}, {Beno{\^\i}t}, {Benoit-L{\'e}vy}, {Bernard},
  {Bersanelli}, {Bielewicz}, {Bikmaev}, {Blanchard}, {Bobin}, {Bock},
  {B{\"o}hringer}, {Bonaldi}, {Bond}, {Borrill}, {Bouchet}, {Bourdin},
  {Bridges}, {Brown}, {Bucher}, {Burenin}, {Burigana}, {Butler}, {Cardoso},
  {Carvalho}, {Catalano}, {Challinor}, {Chamballu}, {Chary}, {Chiang},
  {Chiang}, {Chon}, {Christensen}, {Church}, {Clements}, {Colombi}, {Colombo},
  {Couchot}, {Coulais}, {Crill}, {Curto}, {Cuttaia}, {Da Silva}, {Dahle},
  {Danese}, {Davies}, {Davis}, {de Bernardis}, {de Rosa}, {de Zotti},
  {Delabrouille}, {Delouis}, {D{\'e}mocl{\`e}s}, {D{\'e}sert}, {Dickinson},
  {Diego}, {Dolag}, {Dole}, {Donzelli}, {Dor{\'e}}, {Douspis}, {Dupac},
  {Efstathiou}, {En{\ss}lin}, {Eriksen}, {Finelli}, {Flores-Cacho}, {Forni},
  {Frailis}, {Franceschi}, {Fromenteau}, {Galeotta}, {Ganga},
  {G{\'e}nova-Santos}, {Giard}, {Giardino}, {Giraud-H{\'e}raud},
  {Gonz{\'a}lez-Nuevo}, {G{\'o}rski}, {Gratton}, {Gregorio}, {Gruppuso},
  {Hansen}, {Hanson}, {Harrison}, {Henrot-Versill{\'e}},
  {Hern{\'a}ndez-Monteagudo}, {Herranz}, {Hildebrandt}, {Hivon}, {Hobson},
  {Holmes}, {Hornstrup}, {Hovest}, {Huffenberger}, {Hurier}, {Jaffe}, {Jaffe},
  {Jones}, {Juvela}, {Keih{\"a}nen}, {Keskitalo}, {Khamitov}, {Kisner},
  {Kneissl}, {Knoche}, {Knox}, {Kunz}, {Kurki-Suonio}, {Lagache},
  {L{\"a}hteenm{\"a}ki}, {Lamarre}, {Lasenby}, {Laureijs}, {Lawrence}, {Leahy},
  {Leonardi}, {Le{\'o}n-Tavares}, {Lesgourgues}, {Liddle}, {Liguori}, {Lilje},
  {Linden-V{\o}rnle}, {L{\'o}pez-Caniego}, {Lubin}, {Mac{\'\i}as-P{\'e}rez},
  {Maffei}, {Maino}, {Mandolesi}, {Marcos-Caballero}, {Maris}, {Marshall},
  {Martin}, {Mart{\'\i}nez-Gonz{\'a}lez}, {Masi}, {Matarrese}, {Matthai},
  {Mazzotta}, {Meinhold}, {Melchiorri}, {Melin}, {Mendes}, {Mennella},
  {Migliaccio}, {Mitra}, {Miville-Desch{\^e}nes}, {Moneti}, {Montier},
  {Morgante}, {Mortlock}, {Moss}, {Munshi}, {Naselsky}, {Nati}, {Natoli},
  {Netterfield}, {N{\o}rgaard-Nielsen}, {Noviello}, {Novikov}, {Novikov},
  {Osborne}, {Oxborrow}, {Paci}, {Pagano}, {Pajot}, {Paoletti}, {Partridge},
  {Pasian}, {Patanchon}, {Perdereau}, {Perotto}, {Perrotta}, {Piacentini},
  {Piat}, {Pierpaoli}, {Pietrobon}, {Plaszczynski}, {Pointecouteau}, {Polenta},
  {Ponthieu}, {Popa}, {Poutanen}, {Pratt}, {Pr{\'e}zeau}, {Prunet}, {Puget},
  {Rachen}, {Rebolo}, {Reinecke}, {Remazeilles}, {Renault}, {Ricciardi},
  {Riller}, {Ristorcelli}, {Rocha}, {Roman}, {Rosset}, {Roudier},
  {Rowan-Robinson}, {Rubi{\~n}o-Mart{\'\i}n}, {Rusholme}, {Sandri}, {Santos},
  {Savini}, {Scott}, {Seiffert}, {Shellard}, {Spencer}, {Starck}, {Stolyarov},
  {Stompor}, {Sudiwala}, {Sunyaev}, {Sureau}, {Sutton}, {Suur-Uski}, {Sygnet},
  {Tauber}, {Tavagnacco}, {Terenzi}, {Toffolatti}, {Tomasi}, {Tristram},
  {Tucci}, {Tuovinen}, {T{\"u}rler}, {Umana}, {Valenziano}, {Valiviita}, {Van
  Tent}, {Vielva}, {Villa}, {Vittorio}, {Wade}, {Wandelt}, {Weller}, {White},
  {White}, {Yvon}, {Zacchei}, \& {Zonca}}]{Planck13Cluster}
{Planck Collaboration}, {Ade}, P.~A.~R., {Aghanim}, N., {et~al.} 2014, \aap,
  571, A20, \dodoi{10.1051/0004-6361/201321521}

\bibitem[{{Planck Collaboration} {et~al.}(2016){Planck Collaboration}, {Ade},
  {Aghanim}, {Arnaud}, {Ashdown}, {Aumont}, {Baccigalupi}, {Banday},
  {Barreiro}, {Bartlett}, {Bartolo}, {Battaner}, {Battye}, {Benabed},
  {Beno{\^\i}t}, {Benoit-L{\'e}vy}, {Bernard}, {Bersanelli}, {Bielewicz},
  {Bock}, {Bonaldi}, {Bonavera}, {Bond}, {Borrill}, {Bouchet}, {Bucher},
  {Burigana}, {Butler}, {Calabrese}, {Cardoso}, {Catalano}, {Challinor},
  {Chamballu}, {Chary}, {Chiang}, {Christensen}, {Church}, {Clements},
  {Colombi}, {Colombo}, {Combet}, {Comis}, {Couchot}, {Coulais}, {Crill},
  {Curto}, {Cuttaia}, {Danese}, {Davies}, {Davis}, {de Bernardis}, {de Rosa},
  {de Zotti}, {Delabrouille}, {D{\'e}sert}, {Diego}, {Dolag}, {Dole},
  {Donzelli}, {Dor{\'e}}, {Douspis}, {Ducout}, {Dupac}, {Efstathiou}, {Elsner},
  {En{\ss}lin}, {Eriksen}, {Falgarone}, {Fergusson}, {Finelli}, {Forni},
  {Frailis}, {Fraisse}, {Franceschi}, {Frejsel}, {Galeotta}, {Galli}, {Ganga},
  {Giard}, {Giraud-H{\'e}raud}, {Gjerl{\o}w}, {Gonz{\'a}lez-Nuevo},
  {G{\'o}rski}, {Gratton}, {Gregorio}, {Gruppuso}, {Gudmundsson}, {Hansen},
  {Hanson}, {Harrison}, {Henrot-Versill{\'e}}, {Hern{\'a}ndez-Monteagudo},
  {Herranz}, {Hildebrandt}, {Hivon}, {Hobson}, {Holmes}, {Hornstrup}, {Hovest},
  {Huffenberger}, {Hurier}, {Jaffe}, {Jaffe}, {Jones}, {Juvela},
  {Keih{\"a}nen}, {Keskitalo}, {Kisner}, {Kneissl}, {Knoche}, {Kunz},
  {Kurki-Suonio}, {Lagache}, {L{\"a}hteenm{\"a}ki}, {Lamarre}, {Lasenby},
  {Lattanzi}, {Lawrence}, {Leonardi}, {Lesgourgues}, {Levrier}, {Liguori},
  {Lilje}, {Linden-V{\o}rnle}, {L{\'o}pez-Caniego}, {Lubin},
  {Mac{\'\i}as-P{\'e}rez}, {Maggio}, {Maino}, {Mandolesi}, {Mangilli}, {Maris},
  {Martin}, {Mart{\'\i}nez-Gonz{\'a}lez}, {Masi}, {Matarrese}, {McGehee},
  {Meinhold}, {Melchiorri}, {Melin}, {Mendes}, {Mennella}, {Migliaccio},
  {Mitra}, {Miville-Desch{\^e}nes}, {Moneti}, {Montier}, {Morgante},
  {Mortlock}, {Moss}, {Munshi}, {Murphy}, {Naselsky}, {Nati}, {Natoli},
  {Netterfield}, {N{\o}rgaard-Nielsen}, {Noviello}, {Novikov}, {Novikov},
  {Oxborrow}, {Paci}, {Pagano}, {Pajot}, {Paoletti}, {Partridge}, {Pasian},
  {Patanchon}, {Pearson}, {Perdereau}, {Perotto}, {Perrotta}, {Pettorino},
  {Piacentini}, {Piat}, {Pierpaoli}, {Pietrobon}, {Plaszczynski},
  {Pointecouteau}, {Polenta}, {Popa}, {Pratt}, {Pr{\'e}zeau}, {Prunet},
  {Puget}, {Rachen}, {Rebolo}, {Reinecke}, {Remazeilles}, {Renault}, {Renzi},
  {Ristorcelli}, {Rocha}, {Roman}, {Rosset}, {Rossetti}, {Roudier},
  {Rubi{\~n}o-Mart{\'\i}n}, {Rusholme}, {Sandri}, {Santos}, {Savelainen},
  {Savini}, {Scott}, {Seiffert}, {Shellard}, {Spencer}, {Stolyarov}, {Stompor},
  {Sudiwala}, {Sunyaev}, {Sutton}, {Suur-Uski}, {Sygnet}, {Tauber}, {Terenzi},
  {Toffolatti}, {Tomasi}, {Tristram}, {Tucci}, {Tuovinen}, {T{\"u}rler},
  {Umana}, {Valenziano}, {Valiviita}, {Van Tent}, {Vielva}, {Villa}, {Wade},
  {Wandelt}, {Wehus}, {Weller}, {White}, {Yvon}, {Zacchei}, \&
  {Zonca}}]{Planck15Cluster}
---. 2016, \aap, 594, A24, \dodoi{10.1051/0004-6361/201525833}

\bibitem[{{Rozo} \& {Rykoff}(2014)}]{RozoRykoff14}
{Rozo}, E., \& {Rykoff}, E.~S. 2014, \apj, 783, 80,
  \dodoi{10.1088/0004-637X/783/2/80}

\bibitem[{{Rozo} {et~al.}(2010){Rozo}, {Wechsler}, {Rykoff}, {Annis}, {Becker},
  {Evrard}, {Frieman}, {Hansen}, {Hao}, {Johnston}, {Koester}, {McKay},
  {Sheldon}, \& {Weinberg}}]{Rozo10}
{Rozo}, E., {Wechsler}, R.~H., {Rykoff}, E.~S., {et~al.} 2010, \apj, 708, 645,
  \dodoi{10.1088/0004-637X/708/1/645}

\bibitem[{{Salcedo} {et~al.}(2020){Salcedo}, {Wibking}, {Weinberg}, {Wu},
  {Ferrer}, {Eisenstein}, \& {Pinto}}]{Salcedo20}
{Salcedo}, A.~N., {Wibking}, B.~D., {Weinberg}, D.~H., {et~al.} 2020, \mnras,
  491, 3061, \dodoi{10.1093/mnras/stz2963}

\bibitem[{{Sheldon} {et~al.}(2004){Sheldon}, {Johnston}, {Frieman}, {Scranton},
  {McKay}, {Connolly}, {Budav{\'a}ri}, {Zehavi}, {Bahcall}, {Brinkmann}, \&
  {Fukugita}}]{Sheldon04}
{Sheldon}, E.~S., {Johnston}, D.~E., {Frieman}, J.~A., {et~al.} 2004, \aj, 127,
  2544, \dodoi{10.1086/383293}

\bibitem[{{Shirasaki} {et~al.}(2016){Shirasaki}, {Nagai}, \&
  {Lau}}]{Shirasaki16}
{Shirasaki}, M., {Nagai}, D., \& {Lau}, E.~T. 2016, \mnras, 460, 3913,
  \dodoi{10.1093/mnras/stw1263}

\bibitem[{{Shirasaki} \& {Takada}(2018)}]{Shirasaki18}
{Shirasaki}, M., \& {Takada}, M. 2018, \mnras, 478, 4277,
  \dodoi{10.1093/mnras/sty1327}

\bibitem[{{Simet} {et~al.}(2017){Simet}, {McClintock}, {Mandelbaum}, {Rozo},
  {Rykoff}, {Sheldon}, \& {Wechsler}}]{Simet17}
{Simet}, M., {McClintock}, T., {Mandelbaum}, R., {et~al.} 2017, \mnras, 466,
  3103, \dodoi{10.1093/mnras/stw3250}

\bibitem[{{Spergel} {et~al.}(2015){Spergel}, {Gehrels}, {Baltay}, {Bennett},
  {Breckinridge}, {Donahue}, {Dressler}, {Gaudi}, {Greene}, {Guyon}, {Hirata},
  {Kalirai}, {Kasdin}, {Macintosh}, {Moos}, {Perlmutter}, {Postman},
  {Rauscher}, {Rhodes}, {Wang}, {Weinberg}, {Benford}, {Hudson}, {Jeong},
  {Mellier}, {Traub}, {Yamada}, {Capak}, {Colbert}, {Masters}, {Penny},
  {Savransky}, {Stern}, {Zimmerman}, {Barry}, {Bartusek}, {Carpenter}, {Cheng},
  {Content}, {Dekens}, {Demers}, {Grady}, {Jackson}, {Kuan}, {Kruk}, {Melton},
  {Nemati}, {Parvin}, {Poberezhskiy}, {Peddie}, {Ruffa}, {Wallace}, {Whipple},
  {Wollack}, \& {Zhao}}]{Spergel15}
{Spergel}, D., {Gehrels}, N., {Baltay}, C., {et~al.} 2015, ArXiv e-prints.
\newblock \doarXiv{1503.03757}

\bibitem[{{Stanek} {et~al.}(2010){Stanek}, {Rasia}, {Evrard}, {Pearce}, \&
  {Gazzola}}]{Stanek10}
{Stanek}, R., {Rasia}, E., {Evrard}, A.~E., {Pearce}, F., \& {Gazzola}, L.
  2010, \apj, 715, 1508, \dodoi{10.1088/0004-637X/715/2/1508}

\bibitem[{{Sunayama} {et~al.}(2020){Sunayama}, {Park}, {Takada}, {Kobayashi},
  {Nishimichi}, {Kurita}, {More}, {Oguri}, \& {Osato}}]{Sunayama20}
{Sunayama}, T., {Park}, Y., {Takada}, M., {et~al.} 2020, \mnras, 496, 4468,
  \dodoi{10.1093/mnras/staa1646}

\bibitem[{{Takahashi} {et~al.}(2017){Takahashi}, {Hamana}, {Shirasaki},
  {Namikawa}, {Nishimichi}, {Osato}, \& {Shiroyama}}]{Takahashi17}
{Takahashi}, R., {Hamana}, T., {Shirasaki}, M., {et~al.} 2017, \apj, 850, 24,
  \dodoi{10.3847/1538-4357/aa943d}

\bibitem[{{Tinker} {et~al.}(2008){Tinker}, {Kravtsov}, {Klypin}, {Abazajian},
  {Warren}, {Yepes}, {Gottl{\"o}ber}, \& {Holz}}]{Tinker08MF}
{Tinker}, J.~L., {Kravtsov}, A.~V., {Klypin}, A., {et~al.} 2008, \apj, 688,
  709, \dodoi{10.1086/591439}

\bibitem[{{Tinker} {et~al.}(2010){Tinker}, {Robertson}, {Kravtsov}, {Klypin},
  {Warren}, {Yepes}, \& {Gottl{\"o}ber}}]{Tinker10Bias}
{Tinker}, J.~L., {Robertson}, B.~E., {Kravtsov}, A.~V., {et~al.} 2010, \apj,
  724, 878, \dodoi{10.1088/0004-637X/724/2/878}

\bibitem[{{To} {et~al.}(2020{\natexlab{a}}){To}, {Krause}, {Rozo}, {Wu},
  {Gruen}, {DeRose}, {Rykoff}, {Wechsler}, {Becker}, {Costanzi}, {Eifler},
  {Pereira}, \& {Kokron}}]{To20a}
{To}, C., {Krause}, E., {Rozo}, E., {et~al.} 2020{\natexlab{a}}, arXiv
  e-prints, arXiv:2008.10757.
\newblock \doarXiv{2008.10757}

\bibitem[{{To} {et~al.}(2020{\natexlab{b}}){To}, {Krause}, {Rozo}, {Wu},
  {Gruen}, {Wechsler}, {Eifler}, {Rykoff}, {Costanzi}, {Becker}, {Bernstein},
  {Blazek}, {Bocquet}, {Bridle}, {Cawthon}, {Choi}, {Crocce}, {Davis},
  {DeRose}, {Drlica-Wagner}, {Elvin-Poole}, {Fang}, {Farahi}, {Friedrich},
  {Gatti}, {Gaztanaga}, {Giannantonio}, {Hartley}, {Hoyle}, {Jarvis},
  {MacCrann}, {McClintock}, {Miranda}, {Pereira}, {Park}, {Porredon}, {Prat},
  {Rau}, {Ross}, {Samuroff}, {S{\'a}nchez}, {Sevilla-Noarbe}, {Sheldon},
  {Troxel}, {Varga}, {Vielzeuf}, {Zhang}, {Zuntz}, {Abbott}, {Aguena}, {Annis},
  {Avila}, {Bertin}, {Bhargava}, {Brooks}, {Burke}, {Carnero Rosell}, {Carrasco
  Kind}, {Carretero}, {Chang}, {Conselice}, {da Costa}, {Davis}, {Desai},
  {Diehl}, {Dietrich}, {Everett}, {Evrard}, {Ferrero}, {Flaugher}, {Fosalba},
  {Frieman}, {Garc{\'\i}a-Bellido}, {Gruendl}, {Gutierrez}, {Hinton},
  {Hollowood}, {Huterer}, {James}, {Jeltema}, {Kron}, {Kuehn}, {Kuropatkin},
  {Lima}, {Maia}, {Marshall}, {Menanteau}, {Miquel}, {Morgan}, {Muir}, {Myles},
  {Palmese}, {Paz-Chinch{\'o}n}, {Plazas}, {Romer}, {Roodman}, {Sanchez},
  {Santiago}, {Scarpine}, {Serrano}, {Smith}, {Suchyta}, {Swanson}, {Tarle},
  {Thomas}, {Tucker}, {Weller}, \& {Wester}}]{To20b}
---. 2020{\natexlab{b}}, arXiv e-prints, arXiv:2010.01138.
\newblock \doarXiv{2010.01138}

\bibitem[{{Varga} {et~al.}(2019){Varga}, {DeRose}, {Gruen}, {McClintock},
  {Seitz}, {Rozo}, {Costanzi}, {Hoyle}, {MacCrann}, {Plazas}, {Rykoff},
  {Simet}, {von der Linden}, {Wechsler}, {Annis}, {Avila}, {Bertin}, {Brooks},
  {Buckley-Geer}, {Burke}, {Carnero Rosell}, {Carrasco Kind}, {Carretero},
  {Cunha}, {D'Andrea}, {da Costa}, {De Vicente}, {Desai}, {Diehl}, {Dietrich},
  {Doel}, {Evrard}, {Flaugher}, {Fosalba}, {Frieman}, {Garc{\'\i}a-Bellido},
  {Gaztanaga}, {Gerdes}, {Gruendl}, {Gschwend}, {Gutierrez}, {Hartley},
  {Hollowood}, {Honscheid}, {James}, {Jeltema}, {Kuehn}, {Kuropatkin}, {Lima},
  {Maia}, {March}, {Marshall}, {Melchior}, {Menanteau}, {Miller}, {Miquel},
  {Ogando}, {Romer}, {Sanchez}, {Scarpine}, {Schubnell}, {Serrano},
  {Sevilla-Noarbe}, {Smith}, {Sobreira}, {Suchyta}, {Swanson}, {Tarle},
  {Thomas}, {Tucker}, {Zhang}, \& {DES Collaboration}}]{Varga19}
{Varga}, T.~N., {DeRose}, J., {Gruen}, D., {et~al.} 2019, \mnras, 489, 2511,
  \dodoi{10.1093/mnras/stz2185}

\bibitem[{{Vikhlinin} {et~al.}(2009){Vikhlinin}, {Kravtsov}, {Burenin},
  {Ebeling}, {Forman}, {Hornstrup}, {Jones}, {Murray}, {Nagai}, {Quintana}, \&
  {Voevodkin}}]{Vikhlinin09}
{Vikhlinin}, A., {Kravtsov}, A.~V., {Burenin}, R.~A., {et~al.} 2009, \apj, 692,
  1060, \dodoi{10.1088/0004-637X/692/2/1060}

\bibitem[{{von der Linden} {et~al.}(2014){von der Linden}, {Allen},
  {Applegate}, {Kelly}, {Allen}, {Ebeling}, {Burchat}, {Burke}, {Donovan},
  {Morris}, {Blandford}, {Erben}, \& {Mantz}}]{vonderLinden14WtG}
{von der Linden}, A., {Allen}, M.~T., {Applegate}, D.~E., {et~al.} 2014,
  \mnras, 439, 2, \dodoi{10.1093/mnras/stt1945}

\bibitem[{{Weinberg} {et~al.}(2013){Weinberg}, {Mortonson}, {Eisenstein},
  {Hirata}, {Riess}, \& {Rozo}}]{Weinberg13}
{Weinberg}, D.~H., {Mortonson}, M.~J., {Eisenstein}, D.~J., {et~al.} 2013,
  \physrep, 530, 87, \dodoi{10.1016/j.physrep.2013.05.001}

\bibitem[{{White} {et~al.}(1993){White}, {Navarro}, {Evrard}, \&
  {Frenk}}]{White93}
{White}, S.~D.~M., {Navarro}, J.~F., {Evrard}, A.~E., \& {Frenk}, C.~S. 1993,
  \nat, 366, 429, \dodoi{10.1038/366429a0}

\bibitem[{{Wu} {et~al.}(2015){Wu}, {Evrard}, {Hahn}, {Martizzi}, {Teyssier}, \&
  {Wechsler}}]{Wu15}
{Wu}, H.-Y., {Evrard}, A.~E., {Hahn}, O., {et~al.} 2015, \mnras, 452, 1982,
  \dodoi{10.1093/mnras/stv1434}

\bibitem[{{Wu} {et~al.}(2019){Wu}, {Weinberg}, {Salcedo}, {Wibking}, \&
  {Zu}}]{Wu19}
{Wu}, H.-Y., {Weinberg}, D.~H., {Salcedo}, A.~N., {Wibking}, B.~D., \& {Zu}, Y.
  2019, \mnras, 490, 2606, \dodoi{10.1093/mnras/stz2617}

\bibitem[{{Yu} {et~al.}(2015){Yu}, {Nelson}, \& {Nagai}}]{Yu15}
{Yu}, L., {Nelson}, K., \& {Nagai}, D. 2015, \apj, 807, 12,
  \dodoi{10.1088/0004-637X/807/1/12}

\bibitem[{{Zhang} {et~al.}(2019){Zhang}, {Jeltema}, {Hollowood}, {Everett},
  {Rozo}, {Farahi}, {Bermeo}, {Bhargava}, {Giles}, {Romer}, {Wilkinson},
  {Rykoff}, {Mantz}, {Diehl}, {Evrard}, {Stern}, {Gruen}, {von der Linden},
  {Splettstoesser}, {Chen}, {Costanzi}, {Allen}, {Collins}, {Hilton}, {Klein},
  {Mann}, {Manolopoulou}, {Morris}, {Mayers}, {Sahlen}, {Stott}, {Vergara
  Cervantes}, {Viana}, {Wechsler}, {Allam}, {Avila}, {Bechtol}, {Bertin},
  {Brooks}, {Burke}, {Carnero Rosell}, {Carrasco Kind}, {Carretero},
  {Castander}, {da Costa}, {De Vicente}, {Desai}, {Dietrich}, {Doel},
  {Flaugher}, {Fosalba}, {Frieman}, {Garc{\'\i}a-Bellido}, {Gaztanaga},
  {Gruendl}, {Gschwend}, {Gutierrez}, {Hartley}, {Honscheid}, {Hoyle},
  {Krause}, {Kuehn}, {Kuropatkin}, {Lima}, {Maia}, {Marshall}, {Melchior},
  {Menanteau}, {Miller}, {Miquel}, {Ogando}, {Plazas}, {Sanchez}, {Scarpine},
  {Schindler}, {Serrano}, {Sevilla-Noarbe}, {Smith}, {Soares-Santos},
  {Suchyta}, {Swanson}, {Tarle}, {Thomas}, {Tucker}, {Vikram}, \&
  {Wester}}]{Zhang19}
{Zhang}, Y., {Jeltema}, T., {Hollowood}, D.~L., {et~al.} 2019, \mnras, 1291,
  \dodoi{10.1093/mnras/stz1361}

\end{thebibliography}
\bibliographystyle{aasjournal}

\end{document}